\documentclass[onecolumn,a4paper, showkeys, 10pt]{revtex4}
\usepackage[T1]{fontenc}
\usepackage{amsmath}
\usepackage{amssymb}
\usepackage{graphicx}
\usepackage{csquotes}
\usepackage[colorlinks=true,hyperfootnotes=true,breaklinks=true]{hyperref}

\begin{document}

\title{Modelling Excess Mortality in Covid-19-like Epidemics}

\author{Zdzislaw~Burda}
\email{zdzislaw.burda@agh.edu.pl}

\address{AGH University of Science and Technology, Faculty of Physics and Applied Computer Science, \\ al.~Mickiewicza 30, 30-059 Krakow, Poland; zdzislaw.burda@agh.edu.pl}

\keywords{Epidemic Models; Monte-Carlo Simulations; Random Geometric Networks;
Agent-Based Modelling;}

\begin{abstract}

We develop an agent-based model to assess the cumulative number of deaths during hypothetical Covid-19-like epidemics for various non-pharmaceutical intervention strategies. The~model simulates three interrelated stochastic processes: epidemic spreading, availability of respiratory ventilators and changes in death statistics. We consider local  and non-local modes of disease transmission. The first simulates transmission through social contacts in the vicinity of the place of residence while the second through social contacts in public places: schools, hospitals, airports, {etc.}, where many people meet, who live in remote geographic locations. Epidemic~spreading is modelled as a discrete-time stochastic process on random geometric networks. We use the Monte--Carlo method in the simulations. The~following assumptions are made. The basic reproduction number is $R_0=2.5$ and the infectious period lasts approximately ten days. Infections~lead to severe acute respiratory syndrome in about one percent of cases, 
which are likely to lead to respiratory default and death, 
unless the patient receives an appropriate medical treatment. 
The~healthcare system capacity is simulated by the availability of respiratory ventilators or intensive care beds. Some parameters of the model, like mortality rates or the number of respiratory ventilators per $100 \ 000$ inhabitants, are chosen to simulate the real values for the USA and Poland. In the simulations we compare `do-nothing' strategy with mitigation strategies based on social distancing and reducing social mixing. We study epidemics in the pre-vaccine era, where immunity is obtained only by infection. The model applies only to epidemics for which reinfections are rare and can be neglected. The results of the simulations show that strategies that slow the development of an epidemic too much in the early stages do not significantly reduce the overall number of deaths in the long term, but increase the duration of the epidemic. In particular, a~hybrid strategy where lockdown is held for some time and is then completely released, is inefficient.

\end{abstract}

\maketitle

\section{Introduction \label{sec:intro}}
Mathematical and computer modelling have proved to be very useful tools
for controlling existing infectious diseases \cite{Bailey_1975, Anderson_1992,
Hethcote_2000, Li_2018} as well as for analysing and forecasting epidemics  
\cite{Ferguson_2005, Ferguson_2020, Flaxman_2020}. Modelling~of infectious diseases and epidemics has a long history \cite{Bernoulli_1760, Dietz_1988, Hamer_1906, Ross_1911}. The foundations of the contemporary theoretical epidemiology were laid by W.O. Kermack and A.G. McKendrick \cite{Kermack_1927}. \mbox{Today, theoretical epidemiology} is a mature field of research \cite{Bailey_1975,Anderson_1992, Hethcote_2000, Li_2018}. 

In the last decades, the classical epidemic models 
have been reformulated in the framework of complex networks science 
\cite{Pastor-Satorras_2015}. Complex networks \cite{Barabasi_1999,Albert_2002,Dorogovtsev_2002,Newman_2003} 
are very well-suited to encoding heterogeneity of spatial distribution 
\cite{Barthelemy_2011} and mobility of population 
\cite{Chowell_2003, Colizza_2006, Balcan_2009}. New techniques, which go beyond the classical mean-field approach, have been developed and successfully applied to modelling of epidemic spreading in heterogeneous systems such as degree-based mean-field theory 
\cite{Pastor-Satorras_2001, Barrat_2008}, {models of clustering } \cite{Miller_2009}, 
spatial and mobility networks \cite{Chowell_2003, Colizza_2006, Balcan_2009} {and meta-population} \mbox{approach \cite{Colizza_2007,Colizza_2008}} where one can superimpose hierarchical transportation network on the population distribution in communities, cities, regions and countries, 
to differentiate between disease transmission modes in the regional 
and global scales. The models are based on real-world data and are 
{used to forecast real-world epidemics} \cite{Colizza_2006,Bootsma_2007,Cauchemez_2009,
Bajardi_2011, Otete_2013, Ferguson_2020, Flaxman_2020, Chinazzi_2020, Kucharski_2020,
Zachreson_2019}.

In this article, we are developing a model of a hypothetical epidemic that leads  to Severe Acute Respiratory Syndrome (SARS) for a small fraction of infected people, causing respiratory failure and death. The idea is to mimic some known features of  the Covid-19 epidemic, qualitatively simulate  death statistics during the epidemic and discuss possible control strategies minimising  excess deaths. The~model simulates the spread of epidemic, the availability of respiratory ventilators during epidemic, as well as reference death statistics. When constructing the model, we make the following assumptions:
\begin{enumerate}
\item In the absence of a vaccine, immunity can only be obtained 
through infection.
\item People who get infected become infectious for about ten days.
\item People who recover are immune to reinfection.
\item About one percent of all infections lead to SARS.
\item The occurrence and course of SARS is correlated with the health conditions and age of the infected~person.
\item SARS is likely to lead to respiratory failure and death
unless the person receives appropriate medical attention. 
\item Respiratory ventilation decreases the probability of death. 
\item The death probability is correlated with general health conditions of the patient.
\item The healthcare system has a limited capacity. Especially, the number of doctors and the number of trained medical personnel, and the number of intensive care beds and mechanical ventilators is limited.
\item The mortality rate from non-Covid causes, 
like cancer, cardiovascular diseases, or other chronic diseases, increases during epidemic because of epidemic restrictions in hospitals and health clinics.
\item The epidemic may spread in two distinct modes: via local transmission or non-local (global) transmission.~The local transmission mode corresponds to geographic epidemic spreading through person-to-person contacts near the place of residence.
The non-local transmission mode, in turn, corresponds to epidemic spreading through contacts in public places like: hospitals, cinemas, sport arenas, schools, universities, churches, airports, means of communication, workplaces and many others, where people, who live in different geographic locations, meet. 
\end{enumerate}
We are not attempting to develop a realistic model of Covid-19. Such a model
would have to take into account many detailed medical and demographic factors, 
as well as detailed information about geographical population distribution,  migration, social mixing, {etc.} Instead, we use Occam's razor to develop a model that is as simple as possible and that can be used to qualitatively estimate mortality for a variety of strategies used to restrict epidemic spreading. The idea is to examine dominant factors shaping the death statistics during epidemic whose spread is inhibited by large-scale restrictions on social contacts. We are mainly interested in statistical effects. 

Let us discuss how the above assumptions are implemented in the model.
The numbering below refers to the assumptions.
\begin{enumerate}
\item We are interested in outbreaks like Covid-19,
for which there is initially no immunity or vaccine, and vaccine development
and validation takes several years. We model an epidemic in the pre-vaccine era over 
a time span of $1000$ days, when the only protection mechanism is herd~immunity.
\item The duration of the infection varies from person to person. The model assumes that it is a random variable defined by the geometric law with the mean $\tau = $ 10 days.
The value $\tau=10$ should not be understood 
literally as ten but rather as the order of magnitude. 
The mean incubation period for Covid-19 was estimated to be $5.2\pm 1.8$ {days} \cite{Li_2020}. 
Based on existing literature, the incubation period is $2$ to $14$ days \cite{cdc}. 
We do not distinguish the incubation, latent and infectious periods.  
This simplification does not significantly influence the epidemic dynamics
in large scale, which~is simulated as a variant of {MSIR}  
dynamics \cite{Hethcote_2000}. The rate of epidemic is controlled by the basic reproduction number $R_0$.
According to early estimates \cite{Li_2020,Wu_2020a,Du_2020}, 
the basic reproduction number for Covid-19 ranged from $2.2$ to $2.7$,
while according to an analysis of scientific literature 
on \mbox{Covid-19 \cite{Alimohamadi_2020}}, the mean value of $R_0$ was found to 
range from $1.90$ to $6.49$. We set $R_0=2.5$ as the default value
in the simulations. 
\item The model assumes that a recovered person is immune to the disease or,~alternatively, that~reinfections are rare or do not lead to SARS. 
For some diseases, reinfections are marginal and may be neglected in the description
of epidemic spreading. There is currently an ongoing debate as to whether this is the case with Covid-19 \cite{Ledford_2020}. The model does not apply to epidemics in which
immunity fades over time and reinfections are likely to result in SARS. 
\item The estimated SARS fatality rate for Covid-19 ranges from $0.9\%$ to $2.1\%$ \cite{Wu_2020}.~In the model, the~SARS case frequency rate is $1\%$. This value should be treated as an order of magnitude, \mbox{not a specific} number.
\item It is known that the occurence and course of SARS caused by Covid-19 is correlated with the age and co-existing diseases of the patient. We want to
introduce such correlations to the model in a minimalist way. For this purpose
the agent population is divided into a part where SARS cases are less frequent 
and a part where they occur more frequently. The first part can be thought 
of as healthy or young people, and the second as chronically ill or the elderly.
We label the parts by $H$ and $C$. The parts differ not only by the frequency 
of SARS occurence, but also
by the SARS mortality rates.~The split into the $H$ and $C$ groups may look artificial at first sight but it seems to be the simplest way of implementing the observed variation for Covid-19 of the probability of occurrence and the course of SARS for different age groups. In principle, the~model is well suited to implement many age groups having different patterns for the occurence of Covid-19 SARS and other diseases, but this would make the model more complicated and it would not dramatically change the qualitative picture. So we stick to the simplest solution. We~denote the part sizes $p_H$ and $p_C$, respectively. To fix attention we choose $p_H=75\%$ of the total population and $p_C=25\%$. This split roughly overlaps with the split of the population of a European country (like Poland), into people younger than 60 years ($H$), and older than 60 years ($C$). The~implementation of mortality rates for SARS and other diseases as well as the frequency of SARS occurrence for $H$ and $C$ subpopulations will be presented in detail in the next section where we discuss the stochastic processes that describe the dynamics of the epidemic.
\item The occurence of SARS cases is also simulated as a stochastic process.~An infected person may develop SARS with a certain probability during the infectious period.~In the model, this~probability depends on whether the person is mechanically ventilated or not.  The respiratory ventilation decreases the death probability, so it is important to support by ventilation as many SARS patients as possible.
\item The number of ventilators, or more generally the healthcare system capacity is limited. Not~everyone who needs a support may obtain it in time, when there are too many SARS cases at once. The limited capacity of the healthcare system is simulated in the model by a simple stochastic process of distributing ventilators between SARS patients. Once a patient receives a ventilator he or she will continue to use it until he or she recovers or dies. The ventilator is then transferred to a new person who is randomly selected from all SARS patients who are in need of~one.
\item The death probability is correlated with the general health conditions of the patient. In the model, it is simulated by differentiating the death probability according to the $H$ and $C$ groups to which the patient belongs. The details are given in the next section.
\item The healthcare system capacity is simulated by the number of respiratory ventilators available (or~the number intensive care beds). The number of ventilators differs from country to country. It~is approximately $13.6$ per $100\ 000$ people in Portugal \cite{VPort}, $27.3$ in Russia \cite{VR}, $30.1$~in \mbox{Germany \cite{VG}}, $52.6$ in the USA \cite{VG} and $26.6$ in Poland \cite{VP}. In the study, we use a value $53$, which is close to that for the USA and
$27$ which is close to that for Poland. When assessing excess deaths it is convenient to compare them to the number of deaths when there is no epidemic. The daily deaths per $100\ 000$ people within the period 2007--2014 were $2.74(5)$ in Portugal, $3.78(18)$ in Russia, $2.74(3)$ in Poland, $2.30(18)$ in the USA and $2.48(7)$ in the UK \cite{who}. The numbers in the parentheses correspond to the standard deviations calculated from the eight yearly values in the quoted period. In our study, we use daily deaths' values that simulate those for the US and Poland, {i.e.,} $2.30$ and $2.74$ per $100\ 000$ persons, respectively.
\item The number of deaths from non-Covid causes is expected to increase during the epidemic. The~effect is mainly expected in developed countries where it is related to delayed diagnoses and late admissions of patients with cancer \cite{Richards2020,Maringe2020} and coronary heart diseases \cite{Pessoa-Amorim}.
For example, it was estimated that the diagnosis delays caused by one year of epidemic conditions would lead within 5 years in the UK to an increase of the number of deaths for colorectal cancer by 15.3--16.6\%, for breast 
cancer by 7.9--9.6\%, lung cancer by 4.8--5.3\% and esophageal cancer by \mbox{5.8--6.0\% \cite{Richards2020,Maringe2020}}.~Cancer Research UK has estimated that $2000$ fewer cancers were being diagnosed per week in April 2020 as compared to three years earlier \cite{Richards2020}.
In Poland, in April 2020 only $50\%$ of patients, compared to April 2019, used the system of rapid therapy, which had been introduced some years ago to speed up the treatment of oncological patients. Every third visit to an oncology doctor was canceled, the number of diagnoses using MRI, computed tomography, PET-CT decreased by 
$30\%$ \cite{PP}.

The death risk for people with cardiovascular diseases significantly increases during an epidemic. The number of patients ST-elevation myocardial infarction dropped during the lockdown. More~than $40\%$ of patients with a heart attack were admitted beyond the optimal time \mbox{window \cite{Pessoa-Amorim}}.

These two examples show that protracted epidemic conditions in the healthcare system may have a significant impact on statistics of non-Covid deaths. Some effects will be seen with a time-lag. Cardiovascular diseases and cancer account for the largest share of death statistics. For example there were $647\ 457$ deaths from heart disease, $599\ 108$ from cancer out of a total of $2\ 813\ 503$~deaths in the USA in $2017$ \cite{cdc}. This is  roughly $44.3\%$. Thus, an increase in the number of deaths from these causes by a few percent may be significant for the entire population. People with other chronic conditions will be statistically more exposed to the death risk due to restricted access to healthcare resources during a long-lasting epidemic. Elderly people and people with chronic diseases are fearful of exposure to the virus so they avoid public places including hospitals and health clinics. In effect, they are more exposed to health risks. 
These~phenomena are difficult to model, since they depend on many factors, which cannot be easily quantified, like the organization of the healthcare system, redeployment of resources during epidemic, quarantine procedures in hospitals {\rm etc.}  Instead of seeking a complicated model with many parameters which would describe all these factors we propose to investigate what happens when the rate of deaths, due to causes other than those related to the virus, increases on average by a factor $x$ during the epidemic, where $x$ is just an input parameter of the model. In particular, we study an increase of the daily mortality from non-Covid causes by $x=1\%, 2\%, \ldots, 5\%$. 
\item  Geographic distribution of the population is simulated in the model 
by geometric 2d random networks \cite{Dall_2002, Penrose_2003}, 
see Section \ref{ss_RGN} for details.~Compared to classic random \mbox{graphs \cite{Erdos_1959}}, growing~networks \cite{Barabasi_1999} or other classes of random networks which are constructed in a non-geometric way, such networks are much better at mimicking the distribution of social distances between people in a situation when social contacts in public places and non-local transmissions are limited.~In the model, one can distinguish local and non-local modes of disease transmission.~It is a simplified version of the meta-population \mbox{dynamics \cite{Colizza_2007,Colizza_2008}.}
Local~transmissions are modelled by infections of neighbouring 
nodes of the network, while~non-local ones by infections of randomly 
selected nodes, independently of their position in the network. The non-local mode of
disease transmission simulates intense social contacts in public places
where many people meet, who then move to distant places. 
The effect leads to outbreaks in remote places and therefore significantly accelerates the spread of the epidemic. 
\end{enumerate}

\section{Methods}

In this section we provide a detailed mathematical description of  the model.

\subsection{Random Geometric Networks}
\label{ss_RGN}

Random geometric networks are constructed by the proximity rule \cite{Dall_2002,Penrose_2003}.~Two nodes are connected by an edge if they lie within the given
distance from each other. The simplest example is a network 
constructed by connecting randomly distributed points in a $d$-dimensional 
Euclidean space. We are using this construction here for $d=2$ to mimic geographical distribution of the population  which defines a network of everyday social contacts. For sake of simplicity we assume  
that the points are uniformly distributed on a two-dimensional square 
with the periodic boundary conditions. This can be done by generating 
pairs of coordinates $(x_i,y_i)$, $i=1,\ldots,N$, consisting of $2N$ independent 
random numbers uniformly distributed on the unit interval $[0,1]$
and connecting any two points $i$ and $j$ by an edge of the network 
if the distance between them is smaller than $\epsilon$: 
$\Delta x_{ij}^2 + \Delta y_{ij}^2 \le \epsilon^2$. For the periodic 
boundary conditions the coordinate differences are calculated as follows
$\Delta x_{ij} = \min\left(|x_j-x_i|, 1 - |x_j-x_i|\right)$ and analogously for
$\Delta y_{ij}$. The node degree distribution of the network obtained in this way follows the binomial law
\begin{equation}
P(k) = \binom{N-1}{k} a^k (1-a)^{N-1-k},
\label{Pk}
\end{equation} 
where $a=\pi \epsilon^2$ is the area of a circle of radius $\epsilon$.
The mean degree distribution is $\langle k \rangle = (N-1)a$, and~the variance $\sigma^2(k) = (N-1) a(1-a)$. When $a$ is of the order of $1/N$, then the distribution becomes Poissonian in the large $N$ limit. The node degree distribution (\ref{Pk}) is identical as for  Erd\H{o}s-R\'{e}nyi random graphs \cite{Erdos_1959}. The two classes of graphs are however completely different.
In particular, the average clustering coefficient for the geometric random networks is  
$\langle C \rangle = 1 - \frac{3\sqrt{3}}{4\pi} \approx 0.586503$ \cite{Dall_2002}, while for  Erd\H{o}s--R\'{e}nyi random graphs it approaches zero like $1/N$ as 
$N$ tends to infinity \cite{Erdos_1959}. 

\subsection{Agent-Based Implementation of SIR Dynamics}
\label{ss_SIR}

We use a discrete-time stochastic implementation of the SIR
dynamics \cite{Kermack_1927, Hethcote_2000}. 
The network is populated with 
agents residing on its nodes. The population is divided into three classes of
susceptible (S), infectious (I) and recovered (R) nodes, which describe the 
state of each agent at time $t$. The states change 
in the course of evolution according to epidemic rules which are implemented 
in the model in the form of a discrete time stochastic process. Time is counted
in days from the outbreak of the epidemic. Initially, that is for time $t=0$, 
one agent, or a few ones are infectious, while all others are susceptible. 
An infectious agent remains infective for $\tau$ days on average,
and then it recovers. This~is simulated in the model by assuming that the
probability of remaining infective till the next day is $q$~and of recovering $1-q$. The lifetime distribution
of infectious state is given by a geometric law
\begin{equation}
P_{i}(t) = (1-q) q^{t-1}, \quad t=1,2,\ldots
\label{Pi}
\end{equation}

\noindent The mean lifetime of an infectious state is related to the probability $q$ 
as follows
\begin{equation}
\tau = \langle t \rangle = \sum_{t=1}^\infty t P_{i}(t) = \frac{1}{1-q}
\end{equation}
which means that for 
\begin{equation}
q = \frac{\tau-1}{\tau}
\label{qt}
\end{equation}
the expected infectious period is $\tau$ days. We symbol $\langle \ldots \rangle$
stands for expected value. Clearly,
for $\tau \gg 1$ the probability distribution (\ref{Pi}) can be 
approximated by $P_i(t) \approx e^{-t/\tau}/\tau$. Once an infectious person 
recovers, he or she remains immune and healthy until the end of the SIR of evolution.
Later we will modify the SIR dynamics by superimposing on it the death dynamics
by modifying some of the rules described in this section. In particular, we shall
assume that a recovered person may die with some probability and then reappear as 
a susceptible newborn. This means, in particular, that the $R$ state
may change to $S$ with some probability. We shall discuss the death dynamics in 
the ensuing subsections. The resulting dynamics is similar to that used in 
MSIR models \cite{Hethcote_2000}. 

If an infectious node, $a$, is in contact with a susceptible one, $b$, 
the disease can be transmitted from $a$ to $b$, if the contact is sufficient for
disease transmission. Let $p$ be a probability of transmission from 
$a$ to $b$ in one day. The probability $p_t$ of a transmission within $t$ days 
is $p_t = 1 - (1-p)^t$. The lifetime of an infectious 
state is a random variable (\ref{Pi}) so the transmission probability
for the whole infectious period is equal to the expected value  
$\langle p_t \rangle = 1 - \langle (1-p)^t \rangle$.
This yields 
\begin{equation}
\langle p_t \rangle =  \frac{\tau p}{1 + p(\tau-1)} 
\label{pt_tau}
\end{equation}
for $q$ given by (\ref{qt}). A node has on average $\langle k \rangle$ neighbours, 
so the number of infections generated by a single infected node, in a fully susceptible population, is  
\begin{equation}
R_0 = \langle k \rangle \frac{\tau p}{1 + p(\tau-1)} .
\label{r0p}
\end{equation}

\noindent This equation relates the basic reproduction number $R_0$ 
to the parameters $p$, $\tau$ and $\langle k \rangle$ of the~model. 

The epidemic evolution is implemented in a synchronous way.
This means that all states are updated simultaneously. 
States at time $t+1$ are computed from states at time $t$. 
The following rules are used to update the states. 
If a node is recovered at time $t$, it remains recovered  
at time $t+1$. If a node is infectious at time $t$ it remains infectious
at time $t+1$ with a probability $q$. Otherwise, it changes to recovered.  
If a node is susceptible at $t$ it changes to infectious 
with a probability $p_*$. \mbox{Otherwise, it remains} susceptible.
The probability $p_*$, that a susceptible node becomes infectious is related to the 
transmission probability $p$, by the following relation $p_* = 1- (1-p)^{i_*}$ 
where $i_*$ is an effective number of infectious neighbours
\begin{equation}
i_* = (1-\alpha) i_n + \alpha \frac{\langle k \rangle I}{N},
\end{equation}
and $i_n$ is the number of infectious nearest neighbours of the node
in the network, that is those which
are connected to it by a direct edge. $I$ is the total number of infected nodes in the network. The~parameter $\alpha \in [0,1]$ interpolates between the local and non-local (global) transmission modes.~In the local transmission mode, that is for $\alpha=0$,  
$i_*$ is equal to $i_n$, while in the non-local transmission mode, that is for $\alpha=1$, 
$i_*$ is proportional to all infectious nodes on the network $\langle k \rangle I/N$. 

Later, we shall compare the results of local and non-local transmissions with the results for classic SIR models \cite{Kermack_1927, Hethcote_2000}. In the classic approach one usually uses the continuous time formalism. 
The~epidemic evolution is described by a set of first order 
ordinary differential rate equations for the fractions 
of susceptible, infectious and recovered agents: $s(t)=S(t)/N$, $i(t)=I(t)/N$, $r(t)=R(t)/N$. The epidemic outbreaks if $s(0) R_0>1$. The quantity 
\begin{equation}
\phi(t) = i(t) + s(t) - \frac{1}{R_0} \ln s(t) = \mbox{const} 
\label{phase_p}
\end{equation} 
is conserved during the evolution \cite{Kermack_1927, Hethcote_2000}. 
$s(t)$ is a non-increasing function of time $t$ and $r(t)$ is a non-decreasing function. The infectious fraction, 
$i(t)$ increases for $t< t_{max}$ and reaches a maximum for $t=t_{max}$
such that $R_0 s(t_{max})=1$. Indeed, as one can see from Equation~(\ref{phase_p}), 
the derivative $di/ds = - 1 + \frac{1}{R_0 s}$ changes sign when this condition
is fulfilled. For $t>t_{max}$ the epidemic begins to die out and $i(t)$ decreases from the maximum to zero: 
$i(t) \rightarrow 0$ when $t \rightarrow \infty$.
The fraction of susceptible population for $t\rightarrow \infty$ gives the level
of herd immunity $s(t)\rightarrow s_{hi}$. 
The value $s_{hi}$ can be found from Equation~(\ref{phase_p}).
In particular, if $i(0)$ is very close to zero and $s(0)=1-i(0)$, 
then $s_{hi}$ is a solution to the equation $\ln s_{hi} = R_0(s_{hi} - 1)$.
This yields $s_{hi}\approx 0.4172,\ 0.2032,\ 0.1074,\ 0.0595$ for 
$R_0=1.5,\ 2,\ 2.5,\ 3$, respectively, to give some examples. 

We use the following input parameters in the Monte--Carlo simulations of 
the epidemic on geometric random networks: the number of agents $N$,
the mean node degree $\langle k \rangle$, the basic 
reproduction number $R_0$, the expected duration of the infectious period 
$\tau$, the probability $\alpha$ of long-range transmissions.~As an initial configuration, we choose $I_0$ randomly selected infectious nodes. 
The~remaining nodes are susceptible.   
The probability to remain infectious till the next day is calculated
from Equation (\ref{qt}). The probability of virus transmission
from an infectious to a susceptible agent within one day is calculated from Equation (\ref{r0p}) which gives
\begin{equation}
p = \frac{1}{\tau\left(\frac{\langle k \rangle}{R_0}-1\right) + 1} \ .
\end{equation}

\noindent An example of input values used in the simulations is 
$N=10^5$, $\langle k \rangle=100$, $R_0=2.5$, $\tau=10$, $\alpha=0$,
$I_0=5$. 

\subsection{Modelling Background Conditions}
\label{ss_bg}

In order to assess the impact of epidemics on death statistics, 
one also has to determine the death statistics and the background conditions
in the absence of an epidemic. This is per se an interesting and very complex 
problem since it involves demographic factors, efficiency of healthcare 
systems, statistics of diseases, and many other factors. This is beyond the
scope of this paper. We only model here
basic factors to assess how death statistics change during a pandemic.
The population is divided into classes according to health conditions. In the simplest version of the model we introduce two classes that correspond to healthy people and people with chronic diseases. We label the classes by $H$ and $C$, respectively. The division is symbolic, but it allows the inclusion of statistical correlations between health conditions and mortality in simulations. This is modeled by choosing the mortality rate in the $C$ class to be much larger than in the $H$ class.  The second important difference between the classes is that the death probability during epidemics increases faster in the $C$ class than in the $H$ class. The details are given in the next subsection where we discuss modelling of death statistics.

We assume that the size of the population is constant during the epidemic.
The number of deaths is compensated by the number of newborns. This modifies the SIR dynamics that we described in a simplified version in the previous section. Denote the fraction of healthy people at time $t$ by $h(t)=H(t)/N$, 
the fraction of chronically ill people by $c(t)=C(t)/N$
and the fraction of deaths by $d(t)=D(t)/N$. We have $h(t)+c(t)+d(t)=1$.

We implement the population dynamics as a discrete time stochastic
process (Markov chain) with the following evolution equation
\begin{equation}
\left(h(t+1),c(t+1),d(t+1)\right) = \left(h(t),c(t),d(t)\right)
\left(\begin{array}{ccc} 
p_{HH} & p_{HC} & p_{HD} \\
P_{CH} & p_{CC} & p_{CD} \\
p_{DH} & p_{DC} & p_{DD} \end{array} \right) .
\end{equation}

\noindent The matrix in this equation is a stochastic matrix. It describes
the transition probabilities between the states $H,C,D$. The transition rates
$p_{DH}$ and $p_{DC}$ add up to one
$p_{DH}+p_{DC}=1$, which means that the number of deaths is equal
to the number of newborns. The parameter $p_{DH}$ is the probability that newborns are healthy at birth. For sake of simplicity, but without loss of generality, we additionally assume $p_{HD}=p_{DC}=p_{CH}=0$. The condition $p_{HD}=0$ means that the 
mortality rate of healthy people is zero or it is much smaller than the mortality rate of chronically ill people. The condition $p_{DC}=0$ means that a dead person is replaced by a healthy newborn. Thus the total size of the population is conserved. The condition $p_{CH}=0$ means that a chronically ill person does not become healthy again. Under these assumptions the last equation can be simplified to 
\begin{equation}
\left(h(t+1),c(t+1),d(t+1)\right) = \left(h(t),c(t),d(t)\right)
\left(\begin{array}{ccc} 
1-\beta & \beta & 0 \\
0 & 1-\gamma & \gamma \\
1 & 0 & 0 \end{array} \right) .
\label{transfer}
\end{equation}

\noindent The transfer matrix has only two free parameters: $\beta$---the rate of becoming chronically ill and $\gamma$---the rate of dying. 
This stochastic process has a stationary state
\begin{equation}
\begin{split}
h_* & = \frac{\gamma}{\beta+\gamma+\beta\gamma}, \\
c_* & = \frac{\beta}{\beta+\gamma+\beta\gamma}, \\
d_* & = \frac{\beta\gamma}{\beta+\gamma+\beta\gamma}.
\end{split}
\label{st}
\end{equation}
In our study we choose $\beta$ and $\gamma$ to reproduce the values $d_*=2.3 \cdot 10^{-5}$ or $d_*=2.74 \cdot 10^{-5}$ which correspond to the daily mortality rates in the USA and in Poland, as discussed in Section \ref{sec:intro}. We keep the ratio $h_*/c_*=3$,
so that the simulated population approximately consists of $75\%$ people in the $H$ class and $25\%$ in the $C$ class. For this choice, the paremeters of 
the transfer matrix (\ref{transfer}) are 
\begin{equation} 
\beta = \frac{4}3  \frac{d_*}{1-d_*} \ ,  \gamma = 4 \frac{d_*}{1-d_*},
\label{bgd}
\end{equation}
and $h_*=\frac{3}4 (1-d_*)$ and $c_* = \frac{1}4 (1-d_*)$. 

We conclude this section with two remarks. Firstly, we have assumed that there is no
direct transfer from $H$ to $D$, from $D$ to $C$ and from $C$ to $H$ within
one day, by setting $p_{HD}=0$, $p_{DC}=0$ and $p_{CH}=0$. One~should note
that the probabilities of transfers between these classes in two (or more) days 
are~non-zero
\begin{equation}
\left(\begin{array}{ccc} 
1-\beta & \beta & 0 \\
0 & 1-\gamma & \gamma \\
1 & 0 & 0 \end{array} \right)^2 = 
\left(\begin{array}{ccc} 
(1-\beta)^2 & \beta(2-\beta-\gamma) & \beta\gamma \\
\gamma & (1-\gamma)^2 & \gamma(1-\gamma) \\
1-\beta & \beta & 0 \end{array} \right) \ .
\end{equation}
Secondly, the square or a higher power of the transfer matrix 
(\ref{transfer}) is also a stochastic matrix. In principle, one can replace 
the original transfer matrix with any power of it, and~interpret it as 
a daily transfer matrix. This will not change the stationary state. 
The stationary state is a left eigenvector of the transfer matrix 
associated with the eigenvalue $1$ and it is identical for the transfer matrix (\ref{transfer}) or any power of it. The transfer matrix (\ref{transfer}) has 
three eigenvalues. The one which has the largest absolute value is $\lambda_1=1$ 
and the second largest is $\lambda_2 \approx 1 - \beta -\gamma$. 
The eigenvalue $\lambda_2$ tells us about correlation of states at 
different times $t$, $t'$. The correlation function decays exponentially 
as $\exp(-|t-t'|/T)$. The correlation time $T$ can be derived from 
$\lambda_2$: $T\approx -1/\log(\lambda_2) \approx 1/(\beta+\gamma)$.
For the transfer matrix (\ref{transfer}) $T$ is of order $10^4$. By raising
this matrix to the $n$-th power and interpreting the resultant matrix 
as a daily transfers matrix one can reduce the 
autocorrelation time from $T$ to $T/n$.
  
\subsection{Simulating Death Statistics during Epidemic}
  
Let us begin this section by recalling the philosophy behind splitting the population into parts $H$ and $C$. The mortality rate and the course of SARS for Covid-19  are known to be strongly correlated with age and co-existing diseases. Elderly people and people with chronic conditions die from Covid-19 SARS more frequently than young and healthy people. If one wanted to make the model very realistic one should divide the population into many age groups and, for each, collect good statistics on SARS frequency and mortality and implement these statistics into the model. This would make sense
only if all other elements of the model were realistic.  This is not the case in our study. The model we develop is minimalistic but it should of course implement all important factors, including the correlation between underlying diseases and SARS mortality. The split into two classes with distinct
statistical properties is the simplest way of doing it. For example, we assume
that the frequency, $p_{H,sars}$, of SARS cases in $H$ class is much smaller
than the frequency, $p_{C,sars}$, in the $C$ class. For the sake of simplicity,
we assume that the frequency $p_{C,sars}$ is an order of magnitude larger than
$p_{H,sars}$. The values  $p_{H,sars}$ and $p_{C,sars}$ have to be consistent
with the average SARS frequency which was previously assumed to be $1\%$:
\begin{equation}
p_{sars}= h_* p_{H,sars} + c_* p_{C,sars} \approx  1\%, 
\end{equation}
where $h_*\approx 3/4$ and $c_*=1/4$. In our simulations we use the following values $p_{H,sars}=1/300$ and $p_{C,sars}=3/100$, which give the correct
average.  For this choice, the frequency of SARS cases in the $C$ class
is almost ten times larger than in $H$.  This is the first major difference between $H$ and $C$ classes. Another factor that plays an important role in the death statistics during epidemics is the fatality rate for SARS, which also
should be significantly different for $H$ and $C$. In the model, we
distinguish four situations, labeled by $C_0$, $C_1$, $H_0$, and $H_1$:
\begin{itemize}
\item $C_0$:  Patients with SARS from the $C$ class who are not ventilated; 
\item $C_1$:  Patients with SARS from the $C$ class who are ventilated;
\item $H_0$:  Patients with SARS from the $H$ class who are not ventilated; 
\item $H_1$:  Patients with SARS from the $H$ class who are ventilated.  
\end{itemize}
We assume that the probabilities of dying from SARS are $1.0$, $0.3$, $0.9$ and $0.1$ for $C_0$, $C_1$, $H_0$ and $H_1$, respectively. These values
model a different course of SARS depending on co-existing diseases and access to a ventilator. They mean that respiratory ventilation increases the probability of staying alive from $0\%$ to $70\%$ for people with SARS in the $C$ class, and from $10\%$ to $90\%$ for people with SARS in the $H$ class. 

In the simulations, as input paramaters, we 
use probabilities of dying within one day. They~are related to the probabilities of dying
in the whole period of infection by an equation identical to Equation~(\ref{pt_tau}) in which $\langle p_t \rangle$ is interpreted as the probability of dying from SARS during
the whole period and $p$ is the probability of dying within one day.
For $\tau=10$, the corresponding daily rates are 
$1.0$, $0.041$, $0.474$, $0.011$ for compartments $C_0$, $C_1$, $H_0$, $H_1$, respectively. During an epidemic, the number of people with SARS may easily exceed 
the number of ventilators available. In the simulations we set $V=27$ 
(or $V=53$) ventilators per $100\ 000$ people. These
numbers are close to those for Poland (USA), as~discussed in Section \ref{sec:intro}. A patient with SARS occupies a ventilator until he or she recovers or dies. In~the model, this takes ten days $(\tau=10)$ on average. So, if  for some time there {are more than} $2.7$ ($5.3$) new SARS cases a day per $100\ 000$ people in Poland (USA), the demand for ventilators will exceed the healthcare capacity.  

The ventilator availability is simulated as follows. At any moment of time,
the algorithm keeps track of the number of available ventilators. If this number is larger
than zero, and there is a new SARS case, the number is decreased by one,
and one SARS patient is moved between compartments $C_0$ to $C_1$ or $H_0$ to $H_1$, respectively. The ventilator is occupied until the patient recovers or dies, in which case the number of available ventilators is increased by one.
Initially, the number of ventilators is set to $V$ per $100\ 000$ people.

Another factor that has to be taken into account in assessing the epidemic total death toll is a lower efficiency of the healthcare system during epidemic 
\cite{Richards2020,Maringe2020,Pessoa-Amorim}. This has an impact on the increase of deaths from non-Covid-19 SARS causes. The effect is significant in the group of people with oncological cardiovascular diseases \cite{Richards2020,Maringe2020,Pessoa-Amorim}, but also in the group of people who require continuous medical assistance.
To estimate this effect, systematic statistical surveys should be carried out. Here~we just assume that the number of deaths from other causes than those directly related to SARS increases by a factor $1+x$ during an epidemic, where $x$ is a few percent. In~the model this is implemented by changing the value of the parameter $d_*$ from $d_*$ to $d_*(1+x)$ and recalculating the parameters $\beta$ and $\gamma$ (\ref{bgd}) of the Markov transfer matrix (\ref{transfer}) for days
when the number of infectious agents is $I>0$.

\section{Results}

\subsection{Modes of Infection Transmission}

\begin{figure}[h]
\centering
\includegraphics[width=0.45\textwidth]{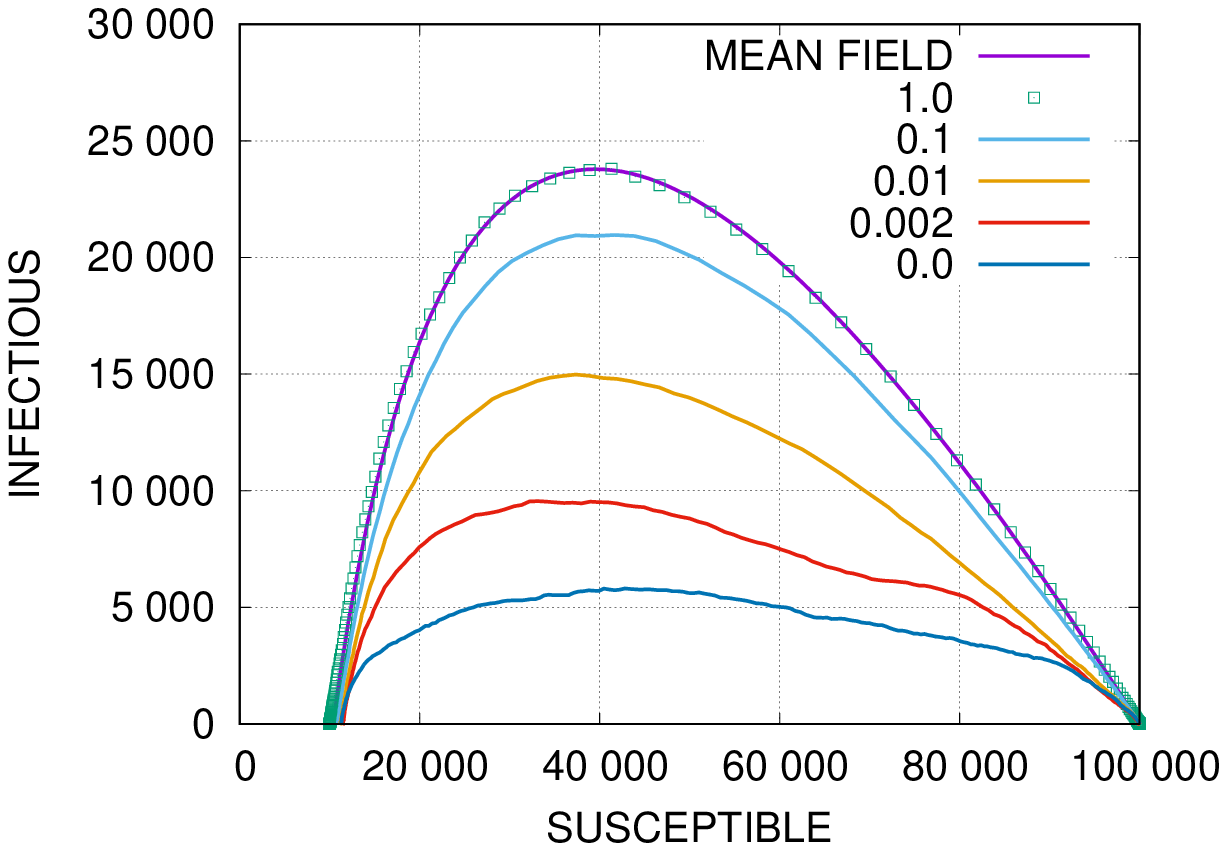}
\includegraphics[width=0.45\textwidth]{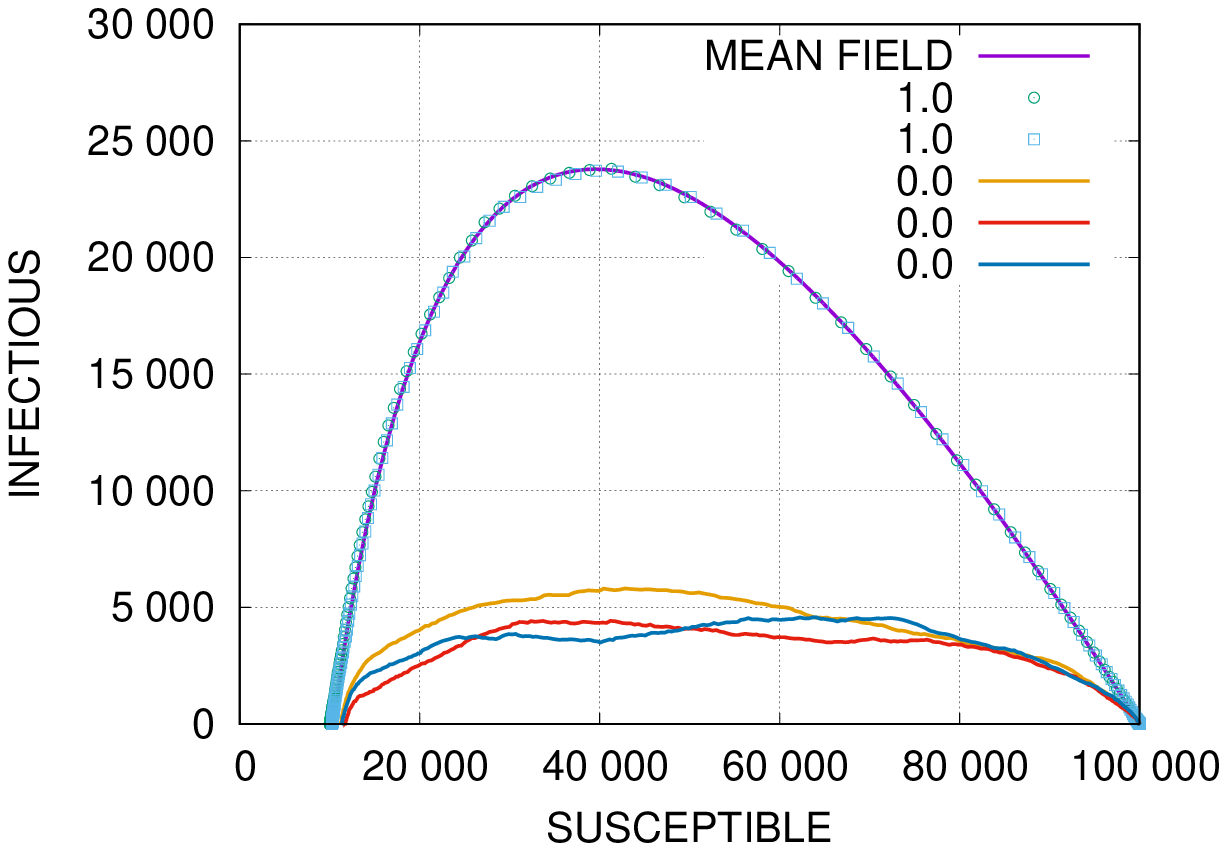}
\caption{\label{fig_fp} (Left) Phase portraits of a simulated epidemic,
with different values of the long-range social mixing parameter $\alpha$, on random geometric random network. The simulations are carried out on networks with $N=10^5$ nodes and the mean node degree $\langle k \rangle =100$. The basic reproduction number used in the simulations is $R_0=2.5$, the infectious period duration is $\tau=10$ and the mixing parameter is $\alpha=1.0,0.1,0.01,0.002,0.0$ (from top to bottom).  The results for $\alpha=1.0$ are shown in symbols and they are compared to a theoretical mean-field result (\ref{phase_p})  (solid line) going through the symbols. The~value of the basic reproduction number in the mean-field result is $R_0=2.53$. (Right) Two different simulations for $\alpha=1.0$ (symbols) compared to the mean-field result (solid line), and three different simulations for $\alpha=0.0$.}
\end{figure}

In the model, the epidemic spreads on a geometric random network through 
local and non-local transmission modes. The non-local mode is selected 
with probability $\alpha$, and the local one with $1-\alpha$, as described before. For $\alpha=1$, the epidemic spreads by the classical SIR mean-field dynamics \cite{Kermack_1927, Hethcote_2000} which depends only on the node degree distribution, 
while $\alpha=0$ it follows a quasi-diffusive dynamics reflecting the geographic population distribution. In Figure \ref{fig_fp} we show phase portraits for epidemics with different values of $\alpha$ on random geometric networks with $N=10^5$ nodes. 
As one can see from Figure \ref{fig_fp} the results of simulations for $\alpha=1.0$ 
are very well described the by the phase-portrait~(\ref{phase_p}) of the classical SIR compartmental model \cite{Kermack_1927, Hethcote_2000}. 
The number of infectious agents it maximal
at $S_{max}/N \approx 1/R_0\approx 0.4$ and the herd immunity is 
achieved for $S_{hi}/N \approx 0.1-0.11$, which~is the place where the curve 
crosses the horizontal axis. This value is close to the mean-field
prediction~(\ref{phase_p}). The~value of the basic reproduction number 
of the best fit to the theoretical curve given by the 
mean-field solution (\ref{phase_p}) is $R_0=2.53$. It differs by one percent from the value $R_0=2.5$ used in the Monte--Carlo simulations. The~difference can be attributed to 
the fact that the classical mean-field dynamics is deterministic 
\cite{Kermack_1927, Hethcote_2000} and $R_0$ is 
a number, while in the simulations the dynamics is stochastic and $R_0$ 
is the mean value of a random variable. The variance of this random 
variable introduces some corrections to the effective value of $R_0$. 

The phase portrait starts to deviate from the mean-field solution when 
$\alpha$ decreases (see Figure \ref{fig_fp}). As shown in the right panel in 
Figure \ref{fig_fp} the phase portraits for different simulations 
for $\alpha=1$ lie on top of each other and are consistent 
with the classical SIR solution. The curves for $\alpha=0$ have stochastic shapes and 
they differ from each other.
 
\begin{figure}[h]
\centering
\includegraphics[width=0.45\textwidth]{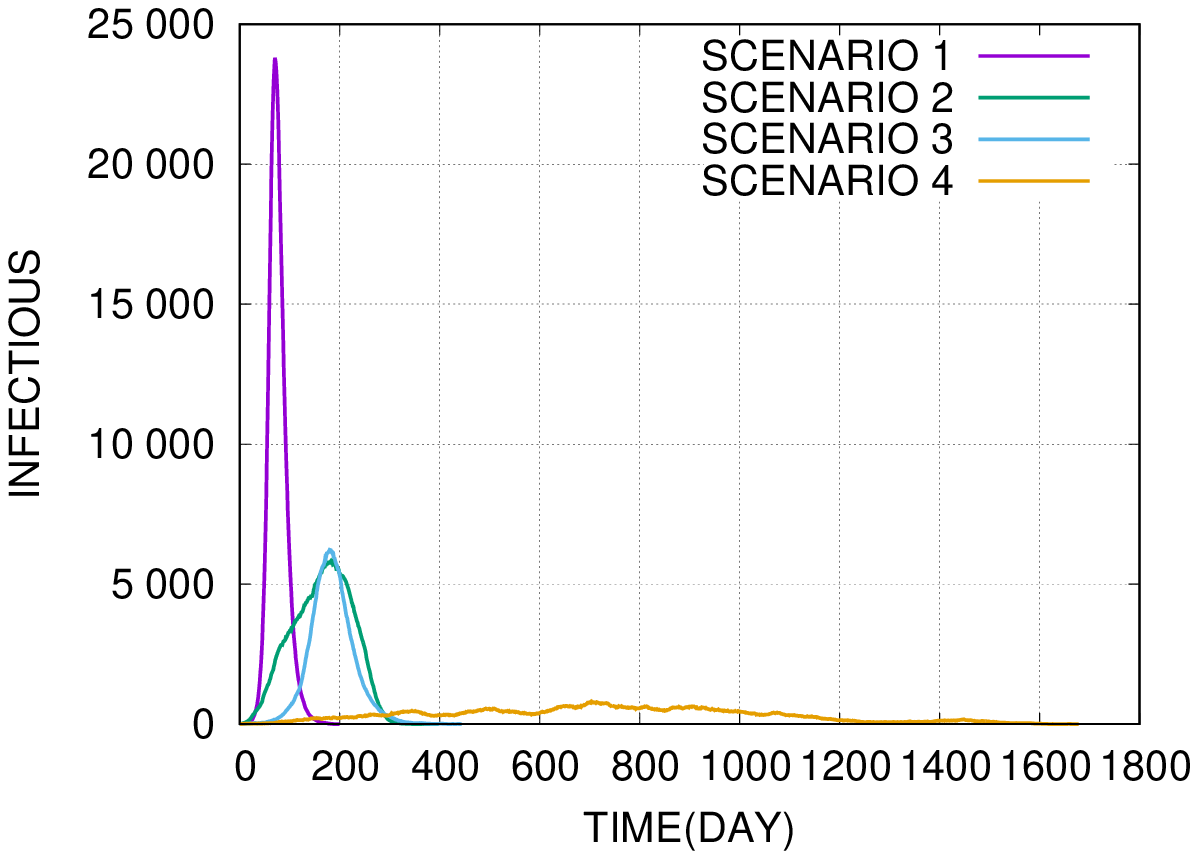}
\includegraphics[width=0.45\textwidth]{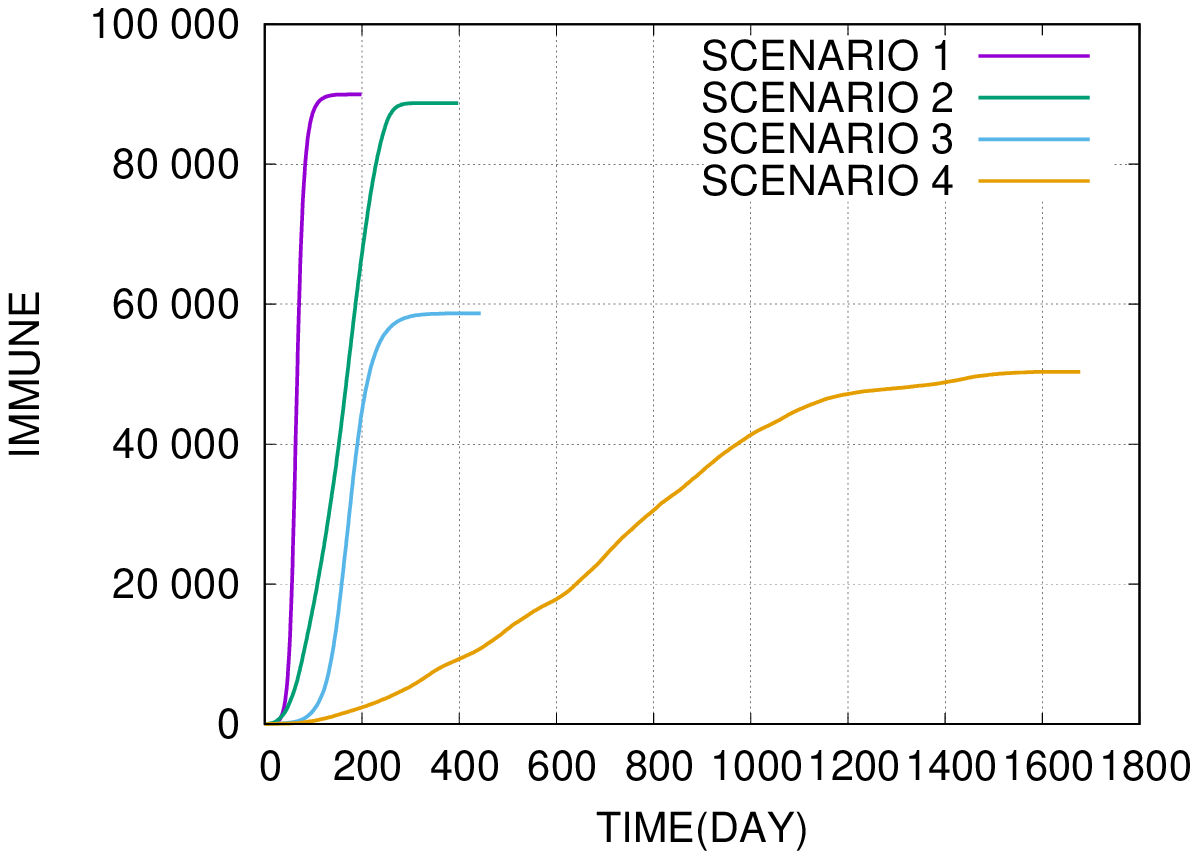}
\caption{\label{fig_SIR}
The charts show the dynamics of epidemic for four scenarios:
(1) $R_0=2.5$, $\alpha=1.0$; (2)~$R_0=2.5$, $\alpha=0.0$; (3) $R_0=1.5$, $\alpha=1.0$;
(4) $R_0=1.5$, $\alpha=0.0$. 
The population size is $N=10^5$, the mean node degree is $\langle k\rangle=100$,
the expected duration of the infectious period is $\tau=10$ in all four cases presented 
in the figures. (Left) The number of infectious agents $I(t)$ as a function of time $t$
expressed in days from the beginning of the epidemic. 
(Right) The number of immune agents: $I(t)+R(t)=N-S(t)$, where $I(t)$, $R(t)$ 
and $S(t)$ are the numbers of infectious, recovered and susceptible agents, respectively.
For scenario 1, the herd immunity level $90\%$ is reached in $t=199$ days.
In scenarios 2, 3, 4, the herd immunity levels: $89\%, 59\%, 50\%$ are 
reached in: $t=398, 444, 1678$ days, respectively.}
\end{figure}

The herd immunity value $S_{hi}$ weakly depends on $\alpha$ (see Figure \ref{fig_fp}).~The~values of \mbox{$S_{hi}/N \approx 0.10-0.11$} are almost identical for $\alpha=1$, and $\alpha=0$.~What depends on $\alpha$ is the
height of the curve which is a few times larger for $\alpha=1$ than 
for $\alpha=0$. This means that long-range social mixing significantly speeds up 
epidemic spreading. The effect is illustrated in Figure \ref{fig_SIR} where  
we compare dynamics of the epidemics for four different scenarios which differ
by the basic reproduction number $R_0$ and the long-range social mixing parameter
$\alpha$. 
One can see that the spread of epidemic depends not only on the basic reproduction 
number $R_0$ but also on the long-range social mixing parameter $\alpha$.
Decreasing the parameter $\alpha$ models closing airports, schools, 
churches, sport arenas, {etc.}, while decreasing the reproduction number $R_0$
models social distancing that is maintaining physical distance between people, 
reducing the frequency of personal contacts, wearing masks as well as 
disinfection, quarantine, isolation, {etc.}~In the next section, we will evaluate
the impact of these measures on mortality during the epidemic
using Monte--Carlo~simulations.

Let us make a couple of remarks to conclude this section.~If one used, in the simulations, Erd\H{o}s--R\'{e}nyi random graphs with exactly the same node degree distribution (\ref{Pk}),~then one would observe the classical mean-field
epidemic dynamics \cite{Kermack_1927, Hethcote_2000} 
 independently of the value of $\alpha$. There~would be no distinction between the local and non-local transmission modes. Spatial~distribution of nodes plays an important role in imitating geographic epidemic spreading. Epidemics spreading in classical random networks \cite{Moore_2020} are completely different than in geometric graphs, or more generally, in~spatial networks, where it has a quasi-diffusive character \cite{Barthelemy_2011}.

The second remark regards the interpretation of results. The trajectories
shown in Figure \ref{fig_SIR} represent single courses of epidemic for the given parameters. The model is stochastic and non-linear so trajectories for other time courses for the same parameters may look differently. For example, the epidemic may die out before it reaches say $1\%$ of the whole population, because of a statistical fluctuation. We performed multiple runs to see how often it ends below the $1\%$ threshold. The results for the four scenarios from Figure \ref{fig_SIR}, are presented in the column $A$ of Table \ref{tab_basic}. The column $T$ shows the average duration time of epidemic, $r_{hi}$ is the immune fraction of the population at the end of the epidemic, and $R_{hi}/T$ is the average number of daily infections. Averages were calculated only from those cases that exceeded the $1 \%$ threshold.

\begin{table}[h]

\centering
\begin{tabular}{ccccc}
\toprule
Scenario & $A$ & $T$ & $r_{hi}$
& $R_{hi}/T$ \\  

\hline

1 & 0.86(22)\%   & 218.5(1.3) & 89.821(12) & 413.6(2.2) \\
2 & 0.91(23)\%   & 420.3(2.8) & 88.768(13) & 212.9(1.4)\\
3 & 12.02(61)\%  & 424.4(3.1) & 59.333(41) & 141.18(98)\\
4 & 14.60(72)\%  & 1529(20)   &  52.05(10) & 35.01(46) \\
  
\hline
\end{tabular}
\caption{Column $A$ shows the percentage of simulated epidemics that expired before reaching $1\%$; $T$ is the average duration of epidemics; $r_{hi}$ is the percentage of recovered people at the end of epidemic; and $R_{hi}/T$ is the average number of new infections per day for the four scenarios shown in Figure \ref{fig_SIR}. \label{tab_basic}}
\end{table}

We shall use the four scenarios in the next section to analyze excess deaths during epidemics. It is, therefore, useful to take a look at the last column of Table \ref{tab_basic} which contains
averages of quotients of $R_{hi}$ at $T$. The values correspond to the average numbers of new infections a day per $100\ 000$ people for scenarios $1,2,3$ and $4$, respectively.
Since the model assumes that the frequency of SARS cases is $1\%$, this means that one can expect c.a. $4.1$, $2.1$, $1.4$ and $0.35$ new SARS cases a day per $100\ 000$ people on average, and much more in the peak. These numbers should be compared with the healthcare system capacity, which is modeled by the number of available ventilators (or ICU beds) which are $V=27$ for Poland and $V=53$ for the USA per $100\ 000$ citizens, as discussed in Section \ref{sec:intro}. A ventilator is occupied on average
for $\tau=10$ days, thus, the maximum capacity of the healthcare system to admit new SARS patients is $V/\tau = 2.7$ or $5.3$.

\subsection{Assessing Mortality Rate for Different Scenarios} 

We are now going to compare excess death statistics for the simulated epidemics 
for six  scenarios: 
\begin{enumerate}
\item $R_0'=R_0=2.5$ and $\alpha=1.0$. This simulates do-nothing strategy. 
An epidemic spreads without any restrictions.
\item $R_0'=R_0=2.5$ and $\alpha=0.0$. This simulates a suppression 
of virus transmission through reducing long-range social mixing.
\item $R_0'=1.5<R_0$ and $\alpha=1.0$. This simulates social distancing
and reduces the transmission rate.  
\item $R_0'=1.5<R_0$ and $\alpha=0.0$. This simulates a quasi-lockdown. Both 
the local and non-local transmission modes are restricted.
\item A quasi-lockdown for $300$ days, as in item 4, and then do-nothing strategy, 
as in item 1.
\item A quasi-lockdown for $600$ days, as in item 4, and then do-nothing strategy, 
as in item 1.
\end{enumerate}

The parameters: $N=10^5$, $\langle k \rangle=100$, $\tau=10$ and $I_0=5$ are 
identical in all simulations.
The six above scenarios are tested in eight systems which differ by
the numbers of ventilators $V$, the daily mortality rates $\mu$, and
the frequency of SARS cases $f$. We consider the following systems:
\begin{enumerate}
\item $\mu = 2.3$; $V=53$; These values are close to the real values for the USA,
so we call the system in short `US'.
\item $\mu = 2.3$; $V=106$;  The number of ventilators is doubled as compared to
that in the USA. We label the system `US-V2'. 
\item $\mu = 2.3$; $V=53$;  The frequency of SARS cases drops from the default
value $f=1\%$ to $f=0.5\%$. This can be interpreted as a consequence of introducing
an effective drug that twice reduces the number of cases of SARS requiring ventilation.
We call this system in short `US-D'.
\item $\mu=2.3$; $V=106$; $f=0.5\%$; This can be interpreted as a result of
doubling the number of ventilators and introducing an effective drug. 
We label the system `US-V2D'.
\end{enumerate}
and four corresponding systems for Poland:
\begin{enumerate}
\setcounter{enumi}{4}
\item $\mu = 2.74$; $V=27$;  $f=1\%$; We call the system in short `PL'.
\item $\mu = 2.74$; $V=54$;  $f=1\%$; We label the system `PL-V2'.
\item $\mu = 2.74$; $V=27$;  $f=0.5\%$; We label the system `PL-D'.
\item $\mu = 2.74$; $V=54$;  $f=0.5\%$; We label the system `PL-V2D'.
\end{enumerate}

As far as the mortality rates and the capacity of the health-care system are concerned the `US' and `PL' systems imitate the situation in the USA and in Poland, while `US-V2' and `PL-V2' simulate a hypothetical situation when 
the capacity of the healthcare systems would have been doubled in the two countries. The  `US-D' and `PL-D' systems, in turn, simulate a situation 
when a pharmaceutic therapy would have reduced the number of SARS patients 
who require mechanical ventilation.  The~resulting effect of introducing an effective drug and doubling the number of ventilators is simulated by the configurations 
`US-V2D' and `PL-V2D'.

The six scenarios in those eight systems are studied for different values of the parameter $x$, which~controls the increase of mortality from 
non-Covid--Sars causes \cite{Richards2020,Maringe2020,Pessoa-Amorim}. 
We scan the range of $x$ from $0\%$ to $5\%$. 
The results are collected in Tables \ref{tab_US}--\ref{tab_PL-V2D}. Each entry corresponds to the average number of additional deaths after $1000$ days per $100\ 000$ people, 
calculated from $100$ independent simulations. 
The values in parenthesis represent statistical uncertainties. 
Only cases of epidemics that exceeded the $ 1\% $ population threshold were included in the analysis. The resulting values should be referred
to the expected number of deaths in $1000$ days per $100\ 000$ people 
in the absence of an epidemic, that is: $ 2740$~in Poland and $ 2300$ in the USA. When analyzing data in the tables, it is worth remembering that 
for scenario 4, the epidemic lasts longer than $1000$ days 
(see Table \ref{tab_basic}).

\begin{table}[h!]

\centering
\begin{tabular}{ccccccc}
\toprule
$x$ & S1 & S2 & S3 & S4 & S5 & S6 \\ 

\hline
0\% & 517.9 (7.8) & 310.3 (5.3) & 136.0 (5.2) & 104.7 (5.7) & 477.9 (6.7) & 352.2 (8.9) \\
1\% & 510.6 (5.9) & 318.7 (5.0) & 156.3 (5.8) & 127.8 (5.8) & 476.8 (7.0) & 336 (10) \\
2\% & 527.1 (5.7) & 330.1 (4.9) & 157.0 (5.1) & 142.1 (5.7) & 484.3 (6.6) & 362 (10) \\
3\% & 523.7 (5.4) & 338.8 (5.3) & 169.5 (5.2) & 172.2 (5.3) & 505.8 (6.4) & 395 (11) \\
4\% & 531.2 (5.5) & 337.1 (4.8) & 185.4 (4.9) & 174.8 (5.4) & 526.8 (6.4) & 417 (11) \\
5\% & 545.6 (5.9) & 359.9 (4.7) & 185.3 (5.0) & 212.3 (5.7) & 540.0 (6.3) & 438 (10) \\

\hline
\end{tabular}
\caption{Excess deaths $1000$ days after the outbreak for `US'. \label{tab_US}}
\end{table}

\begin{table}[h!]
\centering

\begin{tabular}{ccccccc}
\toprule
$x$ & S1 & S2 & S3 & S4 & S5 & S6 \\ 

\hline
0\% & 335.2 (5.6) & 211.2 (4.7) & 144.1 (5.2) & 99.5 (6.3)  & 307.9 (5.7) & 231.6 (6.5) \\
1\% & 340.5 (5.8) & 216.9 (5.3) & 141.0 (5.1) & 140.3 (6.5) & 311.6 (5.9) & 233.7 (4.6) \\
2\% & 340.6 (6.8) & 224.1 (5.1) & 145.3 (4.7) & 153.3 (5.7) & 319.8 (6.4) & 264.0 (6.9) \\
3\% & 347.9 (6.6) & 242.6 (4.6) & 171.9 (4.5) & 160.5 (6.5) & 341.4 (6.9) & 274.7 (5.8) \\
4\% & 355.8 (6.1) & 245.9 (4.5) & 179.2 (5.3) & 196.4 (6.9) & 360.8 (7.3) & 311.9 (7.0) \\
5\% & 357.9 (6.3) & 255.8 (4.7) & 191.5 (5.2) & 218.3 (5.2) & 365.2 (6.7) & 325.7 (7.2) \\

\hline
\end{tabular}
\caption{Excess deaths $1000$ days after the outbreak for `US-V2'. \label{tab:US-V2}}
\end{table}

\begin{table}[h!]

\centering
\begin{tabular}{ccccccc}
\toprule
$x$ & S1 & S2 & S3 & S4 & S5 & S6 \\ 

\hline
0\% & 173.7 (5.1) & 111.0 (4.4) & 169.9 (5.3) & 49.5 (5.8) & 153.8 (5.7) & 114.2 (5.9) \\
1\% & 176.4 (5.0) & 116.1 (4.4) & 176.5 (4.7) & 81.6 (5.0) & 168.4 (5.7) & 130.6 (6.0) \\
2\% & 181.4 (5.8) & 116.9 (4.8) & 186.5 (5.2) & 97.2 (5.2) & 176.7 (5.1) & 162.3 (5.6) \\
3\% & 192.4 (5.7) & 133.9 (4.5) & 199.1 (5.1) & 118.7 (5.7) & 194.8 (5.7) & 170.6 (5.4) \\
4\% & 191.1 (4.8) & 150.2 (5.3) & 208.8 (5.1) & 145.8 (5.2) & 198.6 (5.5) & 190.1 (6.2) \\
5\% & 192.5 (4.9) & 154.5 (4.7) & 216.1 (5.1) & 161.2 (6.4) & 207.5 (5.9) & 214.6 (5.5) \\

\hline
\end{tabular}
\caption{Excess deaths $1000$ days after the outbreak for `US-D'. }

\end{table}

\begin{table}[h!]

\centering
\begin{tabular}{ccccccc}
\toprule
$x$ & S1 & S2 & S3 & S4 & S5 & S6 \\ 

\hline
0\% & 104.6 (4.9) & 111.0 (4.6) & 73.3 (4.9)   & 63.9 (4.3) & 107.6 (5.0) & 110.6 (6.2) \\
1\% & 106.4 (4.7) & 116.5 (5.1) & 71.2 (5.4)   & 69.9 (4.6) & 111.4 (5.4) & 115.9 (5.5) \\
2\% & 121.3 (5.1) & 115.7 (4.5) & 87.5 (5.0)   & 109.4 (6.0) & 127.6 (5.5) & 132.7 (5.3)\\
3\% & 128.1 (5.3) & 133.7 (5.4) & 100.8 (5.2) &  122.4 (5.3) & 138.1 (5.5) & 159.3 (5.0)\\
4\% & 120.4 (4.6) & 133.7 (5.1) & 112.2 (5.0) &  141.4 (5.2) & 154.6 (5.7) & 177.1 (4.8)\\
5\% & 132.5 (5.2) & 159.3 (5.4) & 119.4 (5.9) & 155.7 (6.4) & 162.5 (5.2) & 193.4 (5.2) \\

\hline
\end{tabular}
\caption{Excess deaths $1000$ days after the outbreak for `US-V2D'.}
\end{table}

\begin{table}[h!]

\centering
\begin{tabular}{ccccccc}
\toprule
$x$ & S1 & S2 & S3 & S4 & S5 & S6 \\ 

\hline
0\% & 617.9 (7.0) & 290.1 (6.4) & 225.6 (5.4) & 93.9 (7.1) & 596.8 (7.1) & 460 (12)  \\
1\% & 628.5 (5.9) & 294.3 (6.6) & 242.2 (6.0) & 127.6 (6.7) & 595.1 (7.1) & 477 (14) \\
2\% & 632.6 (6.3) & 304.1 (7.5) & 257.0 (6.3) & 163.1 (5.9) & 617.7 (7.8) & 491 (13)  \\
3\% & 634.0 (6.1) & 323.4 (6.0) & 266.1 (5.0) & 186.5 (6.9) & 630.2 (7.4) & 519 (13) \\
4\% & 642.6 (6.0) & 333.7 (6.4) & 277.0 (6.5) & 208.4 (6.6) & 638.2 (6.7) & 544 (11) \\
5\% & 648.7 (6.5) & 340.3 (6.2) & 292.4 (6.1)  & 220.3 (7.6) & 645.5 (6.8) & 567 (11) \\  

\hline
\end{tabular}
\caption{Excess deaths $1000$ days after the outbreak for `PL'.}
\end{table}

\begin{table}[h!]

\centering
\begin{tabular}{ccccccc}
\toprule
$x$ & S1 & S2 & S3 & S4 & S5 & S6 \\ 

\hline
0\% &  498.9 (6.1) & 207.4 (5.8) & 143.5 (5.5) & 91.5 (6.9) & 462.9 (6.1) & 342 (10) \\
1\% &  508.6 (6.1) & 225.7 (5.1) & 148.5 (5.8) & 127.9 (7.1) & 487.5 (7.1) & 366 (10) \\
2\% &  512.4 (6.0) & 232.8 (5.7) & 163.7 (5.3) & 149.8 (5.5) & 486.0 (7.6) & 379 (12) \\
3\% &  513.6 (5.1) & 244.2 (5.7) & 174.8 (5.4) & 187.9 (5.6) & 504.6 (7.2) &  383 (10)  \\
4\% &  529.5 (5.9) & 249.8 (5.5) & 184.6 (5.2) & 199.5 (7.0) & 520.7 (7.2) & 404 (11) \\
5\% &  529.5 (5.8)  & 269.1 (6.4) & 194.0 (6.2) & 230.2 (4.9) & 524.9 (6.0) & 442 (11) \\

\hline
\end{tabular}
\caption{Excess deaths $1000$ days after the outbreak for `PL-V2'. \label{tab:PL-V2}}
\end{table}

\begin{table}[h!]

\centering
\begin{tabular}{ccccccc}
\toprule
$x$ & S1 & S2 & S3 & S4 & S5 & S6 \\

\hline
0\% & 262.1 (5.5) & 98.9 (5.8) & 76.0 (5.9) & 42.5 (5.8) & 238.1 (6.2) & 163.4 (6.5) \\
1\% & 261.8 (6.0) & 113.6 (5.3) & 87.9 (6.2) & 65.7 (5.6) & 247.8 (7.7) & 185.5 (7.1) \\
2\% & 269.3 (5.5) & 123.2 (5.5) & 84.3 (5.7) & 101.4 (5.2) & 265.2 (6.1) & 215.2 (6.7) \\
3\% & 272.6 (5.1) & 139.0 (5.5) & 109.8 (5.1) & 123.3 (5.6) & 268.0 (5.6) & 231.9 (7.2) \\
4\% & 278.4 (5.2) & 158.4 (5.0) & 124.4 (5.8) & 157.5 (5.4) & 290.1 (7.0) & 260.4 (7.4) \\
5\% & 282.7 (5.0) & 159.5 (4.7) & 134.8 (5.9) & 189.7 (5.9) & 304.3 (5.9) & 280.0 (7.2) \\

\hline
\end{tabular}
\caption{Excess deaths $1000$ days after the outbreak for `PL-D'.}
\end{table}

\begin{table}[h!]

\centering
\begin{tabular}{ccccccc}
\toprule
$x$ & S1 & S2 & S3 & S4 & S5 & S6 \\

\hline

0\% & 163.3 (5.9) & 103.0 (5.6) & 66.7 (5.5) &  59.2 (5.7) & 145.0 (6.3) & 112.6 (6.6) \\
1\% & 162.2 (5.5) & 111.4 (5.4) & 82.5 (5.8) &  81.4 (6.3) & 166.2 (6.3) & 129.2 (5.2) \\
2\% & 180.7 (6.2) & 127.9 (4.9) & 90.4 (5.6) &  104.7 (5.8) & 182.2 (5.3) & 154.0 (6.3) \\
3\% & 191.1 (5.6) & 134.5 (5.2) & 110.5 (5.7) & 126.7 (6.1) & 204.9 (6.5) & 178.5 (6.4) \\
4\% & 188.7 (5.4) & 146.0 (5.5) & 112.5 (5.1) & 155.9 (6.1) & 199.7 (6.1) & 199.0 (6.1)\\
5\% & 199.8 (5.3) & 158.8 (5.2) & 126.0 (5.5) & 184.5 (6.3) & 209.6 (7.1) & 224.6 (6.1)\\

\hline

\end{tabular}
\caption{Excess deaths $1000$ days after the outbreak for `PL-V2D'.\label{tab_PL-V2D}}
\end{table} 

\begin{figure}[h]
\centering
\includegraphics[width=0.45\textwidth]{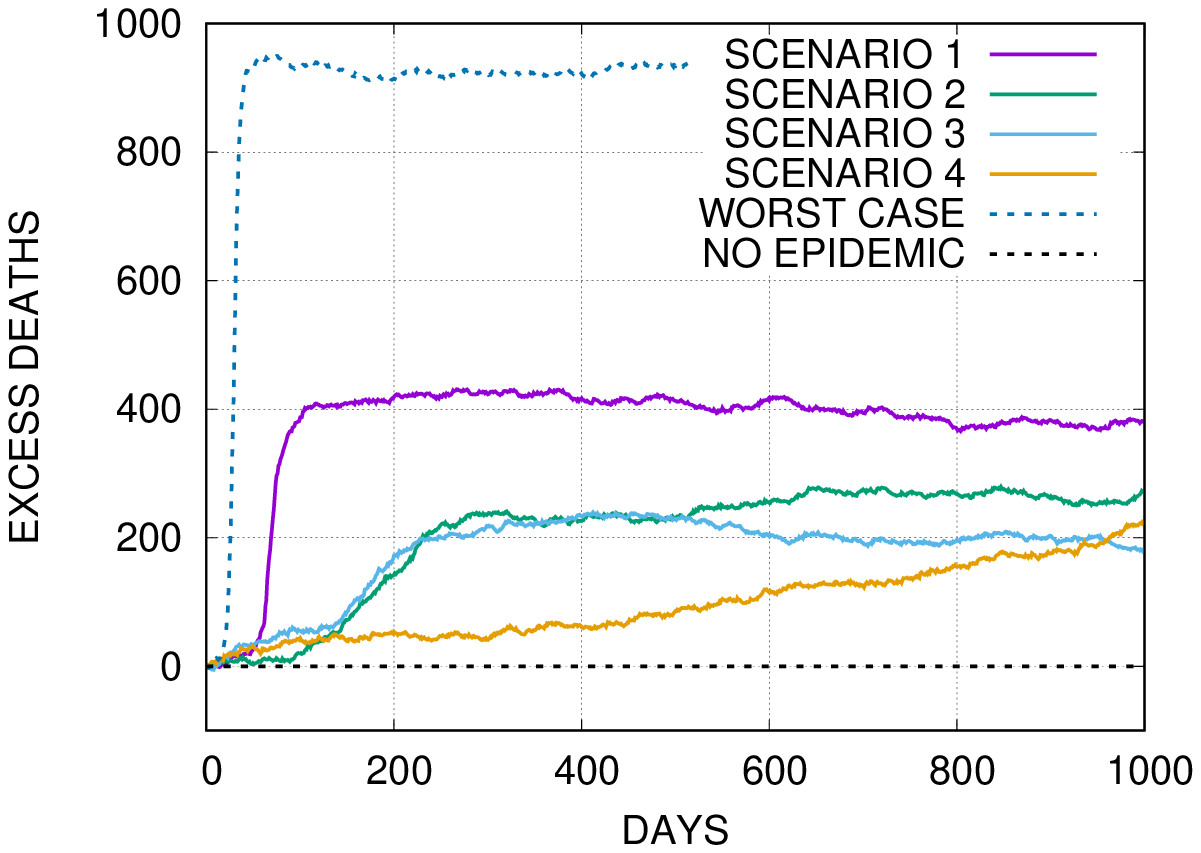}
\includegraphics[width=0.45\textwidth]{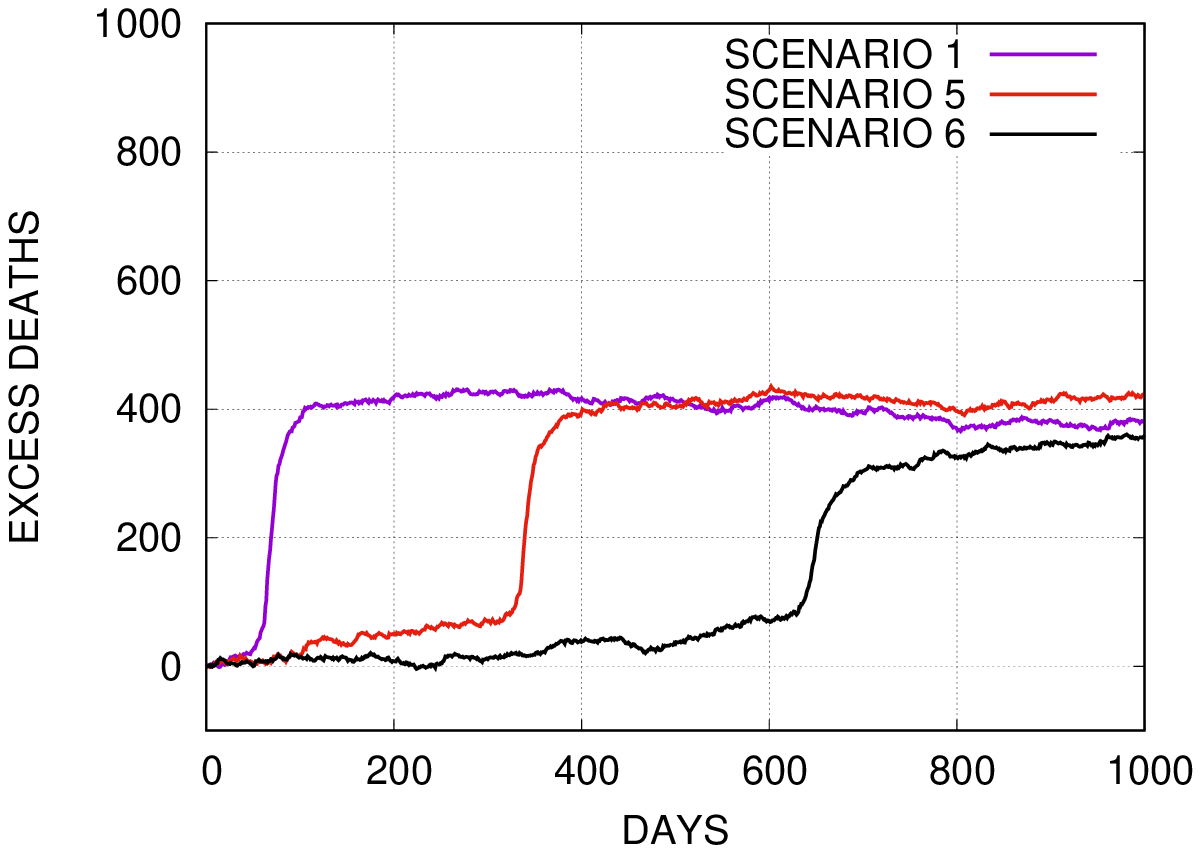}
\caption{\label{fig_tdt} 
The cumulative number of excessive deaths for different scenarios in Monte--Carlo 
simulations of an epidemic for a population of $N=10^5$ agents on a random geometric network for the system `US-V2' and for $x=0.05$. (Left) The upper curve corresponds to the worst-case scenario, that is, none of the SARS patients receive medical attention during the epidemic.
The four curves below correspond to the scenarios 1-4 presented 
in the main text and in Figure~\ref{fig_SIR}. The dashed line in the bottom 
represents the background mortality. (Right) The graphs show the
cumulative number of excessive deaths for scenarios 5 and 6. For reference, also the curve for scenario 1, which is identical as in the left panel, is shown.}
\end{figure}

Let us, for illustration, present some results graphically. In Figure \ref{fig_tdt} we show an example of time evolution of the number of additional deaths during $1000$ days after the outbreak in the system `US-V2' for six different  scenarios. 

The first four scenarios are shown in the left panel in Figure \ref{fig_tdt} 
and the remaining two in the right one. In the left panel, we additionally draw two reference curves representing the worse-case scenario when no SARS patients receive required medical assistance during an epidemic, and the scenario when there is no epidemic. For scenario 1, the number of daily new infections is large and the number of SARS cases exceeds the healthcare system capacity,
so the number of daily SARS deaths is large. The~effect manifest as a steep part
of the mortality curve. The epidemic lasts a short time. \mbox{For scenarios 2, 3} the epidemic lasts longer but the number of daily new infections is much lower. In~effect, most of SARS patients obtain required medical attention, so the daily excess  mortality rate is much lower than in scenario 4. In scenario 4, the epidemic spreads very slowly. The number of the new daily SARS cases is small, much below
the healthcare system capacity. SARS patients are optimally treated,
however, deaths from causes other than Covid are increasing due to
protracted epidemiological restrictions.
The graphs in the right figure show what happens when the lockdown lasts for
$300$ or $600$ and then it is completely lifted.
One can see that at the end of the studied period the total number of deaths is roughly the same as in the `do-nothing' strategy, shown in the figure for~reference.

\begin{figure}[h]
\centering
\includegraphics[width=0.45\textwidth]{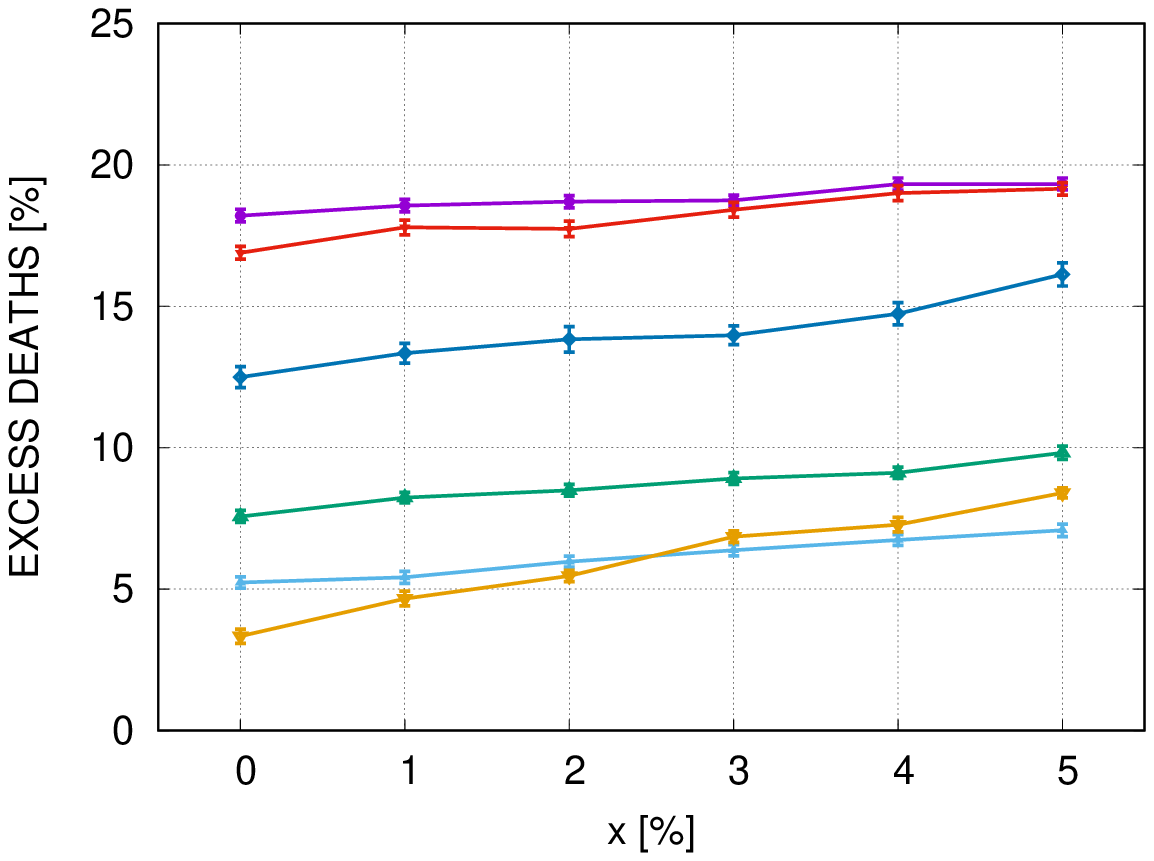}
\includegraphics[width=0.45\textwidth]{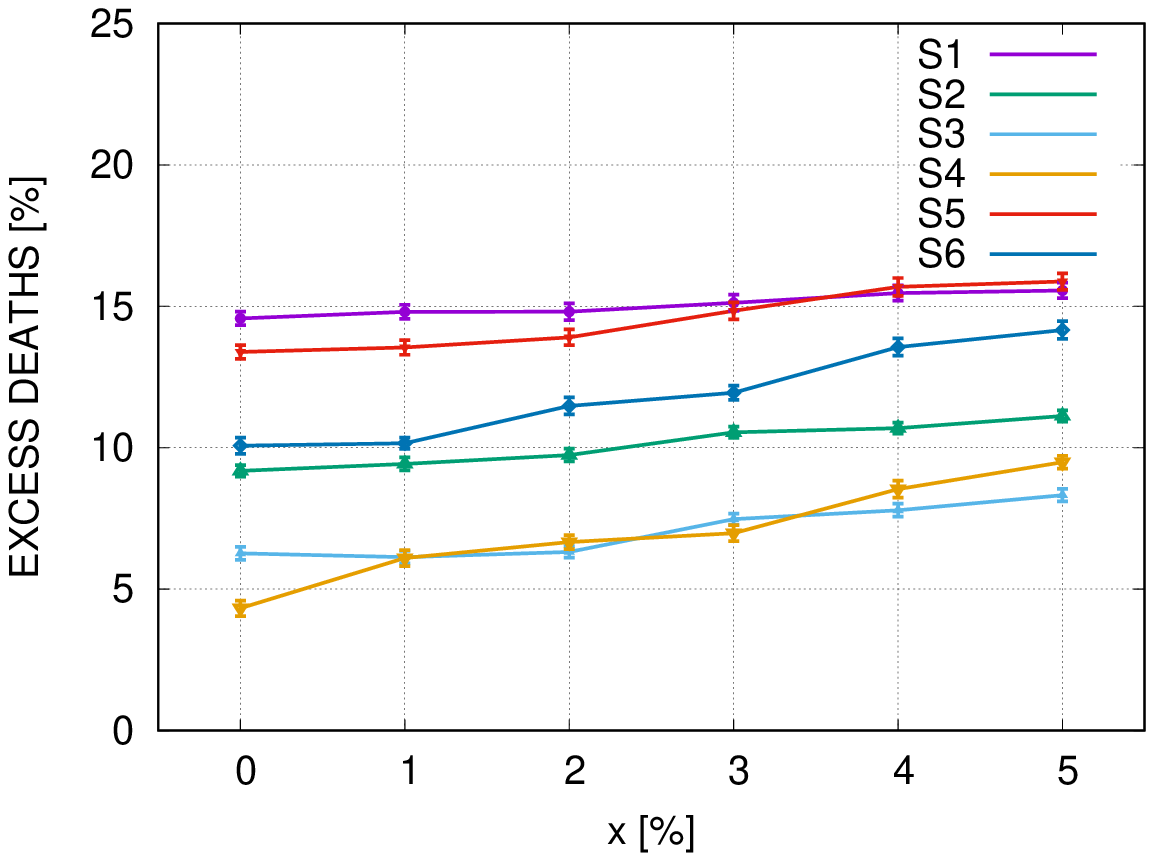}
\caption{\label{fig_V2} The number of excessive deaths relative to the expected
number deaths in the absence of epidemic, in the PL-V2 system (Left) and US-V2 (Right), for six scenarios described in the main text. The points with error bars represent values from Tables \ref{tab:PL-V2} and \ref{tab:US-V2} divided by $2740$ and $2300$, respectively. Lines between the points are drawn to guide the eye.} 
\end{figure}

\begin{figure}[h]
\centering
\includegraphics[width=0.45\textwidth]{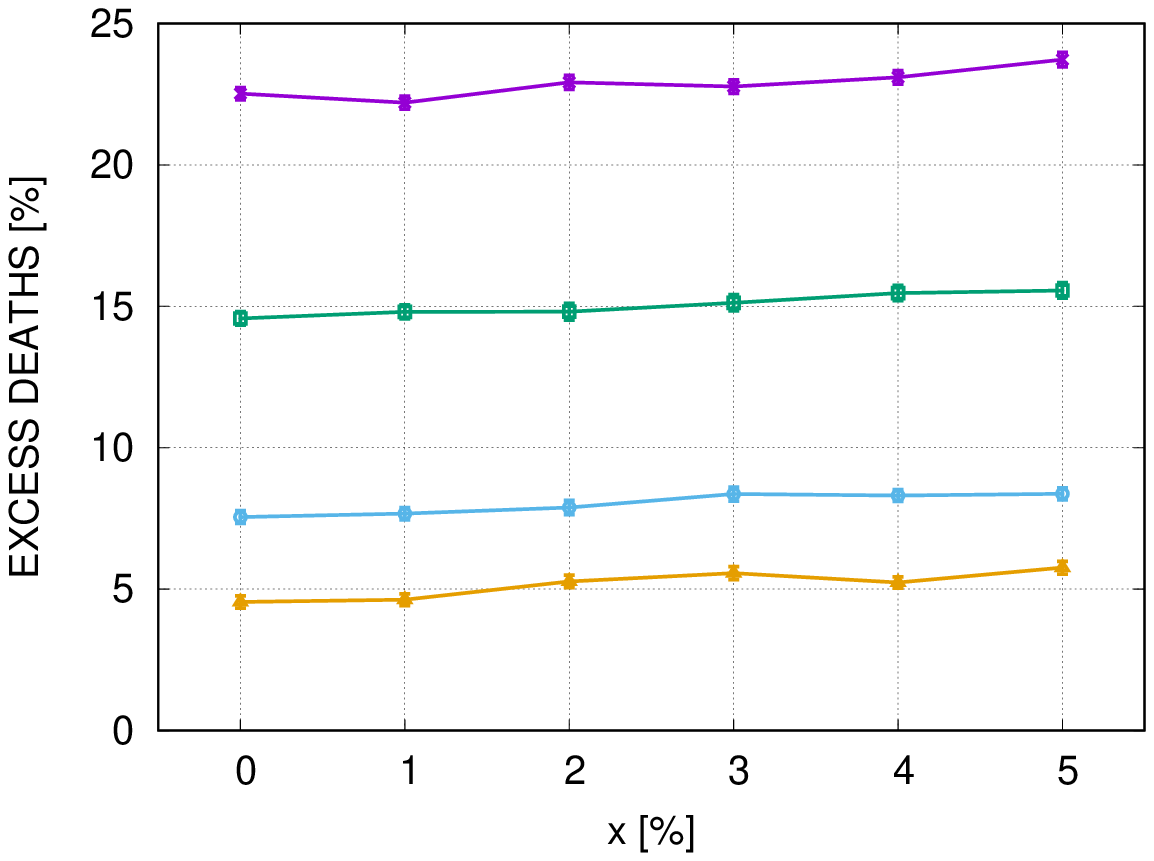}
\includegraphics[width=0.45\textwidth]{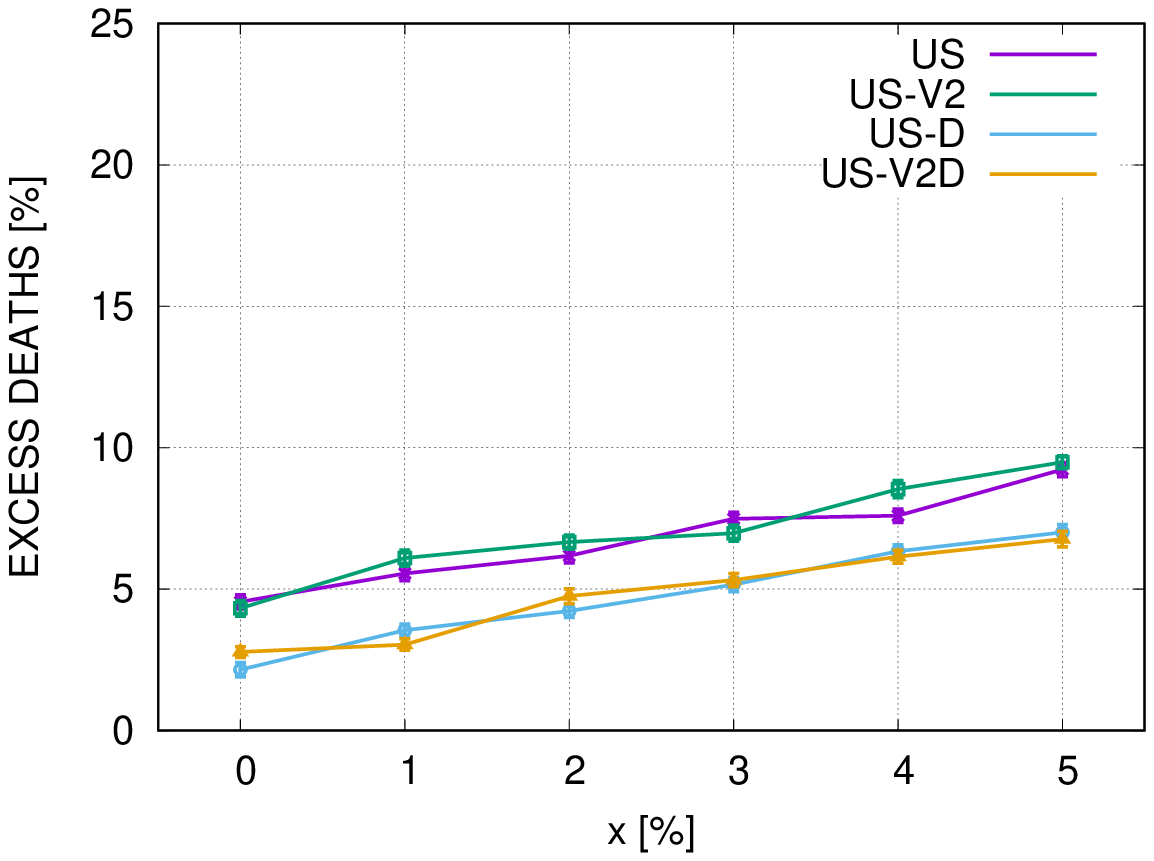}
\caption{\label{fig_S1_S4} Comparison of strategy 1 (Left) and strategy 4 (Right)
for the configuration `US', `US-V2', `US-D' and `US-V2D'.}
\end{figure}

As the next example, we compare in Figure \ref{fig_V2}, additional deaths
$1000$ days after the outbreak for all six scenarios in the systems `PL-V2' and `US-V2' which simulate hypothetical situations of the doubled capacity of the healthcare systems in Poland and the USA.  The slope of the graphs increases with the duration of the epidemic as additional deaths from causes other than Covid are increasing with time. In particular one can see that the curves for scenarios 3 and 4 intersect for $x$ close to $2\%$. In other words, these two strategies are comparable in this case. For economic reasons, strategy 3 is, however, much better than strategy 4, because it takes much less time. We also see in the figure that strategies 1 and 6 lead roughly to the same number of additional deaths.
In Figure \ref{fig_S1_S4} we compare the effect of doubling the healthcare system
capacity, measured by the number of ventilators (or ICU beds) in scenario 1
(`do nothing strategy') and scenario 4 (`quasi-lockdown') in the USA. 
We use parameters as for the `US', `US-V2', `US-D' and `US-V2D' systems. In the left figure, we plot graphs for the `do-nothing strategy'. We see that excessive mortality is approximately $23\%$ and slowly varies with $x$. If
the number of ventilators doubled, the excessive mortality would drop to around $15\%$. The introduction of a drug, that reduces the number of SARS cases requiring respiratory ventilation to $0.5 \% $, would reduce the excessive mortality to approximately $8\%$, and if additionally, the number of ventilators doubled, to approximately $5\%$.  The picture
is completely different for scenario 4, as shown in the right figure. We see that
the graphs for `US' and `US-V2' basically overlap, meaning that 
the doubling of the number of ventilators has no effect on mortality in this case.  The same holds for `US-D' and `US-V2D'. Clearly, in scenario 4, the quantities of daily infections are so small that the healthcare system has a sufficient capacity. Additional ventilators are unnecessary in this case.

\section{Discussion}

We conducted a Monte--Carlo study of epidemic spreading on random geometric
networks to assess the efficiency of non-pharmaceutic interventions in reducing 
the total number of surplus deaths during Covid-19-like epidemics.
We discussed strategies based on social distancing and restricting 
long-range social mixing. They have different effects on epidemic
spreading. Social distancing reduces the basic reproduction
number $R_0$ to some effective reproduction number $R_0'<R_0$. Restrictions
on long-range social mixing reduce virus transmission between remote places. 
When long-range mixing is large, an epidemic spreads
via mean-field dynamics. When it is small, it spreads via quasi-diffusive 
dynamics depending on the geographic population distribution. We studied
the two modes of disease transmission here.

There are two sources of deaths which contribute to the total death toll
during a Covid-19-like epidemic. One is related to Covid-SARS,
and the other one to other diseases. The number of SARS deaths depends on the capacity of the healthcare system, which in the model is simulated by the number of available respiratory ventilators. If the daily number of new SARS cases exceeds $V/\tau$, where~$V$ is the number of ventilators and $\tau$ is the number of days of using one ventilator for one SARS patient, some people with SARS will not be ventilated and will have lower survival probability. This~effect was simulated in the model. If one assumes
that there are $27$ or $53$ ventilators per $100\ 000$~people, as in Poland
or the USA, and $\tau$ is approximately $10$ days, then $V/\tau=2.7$ or
$5.3$. As long as the number of SARS patients is below $V/\tau$, then the number of deaths caused by SARS is maximally reduced. This effect can be achieved by slowing down the epidemic. On the other hand, the number of excess deaths from other causes may increase \cite{Richards2020,Maringe2020,Pessoa-Amorim} with the epidemic duration so it is not beneficial to slow the rate of epidemic spreading too much. The optimal solution is to keep the number of SARS cases close to the capacity of the healthcare system, but not much below it. 

We also showed that a strategy of maintaining the lockdown for some time and then releasing it by removing all restrictions has a similar effect on
the number of deaths in the long term as if the do-nothing strategy was introduced right 
at the beginning. The deaths differ only by the time when they occur: 
in the do-nothing strategy the mortality is large at the beginning  
while in the other case it is large when the lockdown is released. 

A strict lockdown makes sense only when one wants to gain time to increase the healthcare system capacity, for instance, buying new ventilators, increasing the number of ICU beds, training medical personnel or improving medical and epidemic procedures, 
or when an effective drug or vaccine is expected to be introduced in a short time.  Otherwise, the optimal strategy is to keep the epidemic progress at the level that the number of SARS patients at any time is roughly equal to the capacity of the healthcare system. If the number of SARS cases is much larger than that, too many people will die of SARS. If it is much smaller than that and the epidemic will last too long, many additional people can die from cancer, cardiovascular diseases and chronic diseases, due to later diagnoses, later admissions for hospitalization and restricted access to health services \cite{Richards2020,Maringe2020,Pessoa-Amorim}.

Social distancing reduces the herd immunity level, see Table \ref{tab_basic}.
This means that after lifting restrictions on social distance and restoring normal contacts between people, the percentage of immune people will be below the herd immunity level for the unrestricted system. The system will be unstable, in the sense that a new single infectious person may trigger a new outbreak. The situation is similar to a superheated liquid, where boiling may occur spontaneously at any time. For example, the total number of deaths in the simulated epidemic is comparable for scenarios 2 and 3 (see Figure \ref{fig_tdt}), but the percentage of immune people at the end of the epidemic is $89\%$ in scenario 2 and $59\%$ in scenario 3 (see~Table~\ref{tab_basic}). The value $89\%$ is close to the herd immunity level for $R_0=2.5$ while $59\%$ is far below it. This means that the epidemic in scenario 3 can restart from the level $59\%$ when the restrictions are lifted and a new infective person appears. This example shows that strategies reducing long-range social mixing bring better effects than introducing social distancing locally. They are, however, much~more difficult to implement.

Let us compare the current Covid-19 mortality to typical mortality rates in Poland.
Rates are quoted as daily deaths per $100\ 000$ people. 
For example, in $2014$ the daily mortality from all causes was approximately $2.71$ including $0.69$ from cancer \cite{who,krn}. The daily number of deaths registered as Covid-19 deaths between March the $5$th and September the $27$th, $2020$, in Poland was approximately $0.031$ \cite{who_covid}. According to the WHO data, the cumulative number of registered Covid infections in Poland in the quoted period was $227.2$ per $100\ 000$ people (approximately $2$ per mille). The rate of spreading for Covid-19
is much slower than the simulated epidemics (see Table \ref{tab_basic}). Covid-19 has been spreading very slowly so far. If Covid-19 continued spreading at this rate the epidemic would need many years to end, unless an efficient vaccine is introduced. The number of registered cases is probably much smaller than the real number of Covid-19 cases because mainly suspected cases have been tested so far, so the statistics may be very biased. Assuming the number of cases is up to ten times underestimated, that would mean that at the end of September $2020$ approximately $2\%$ of people in Poland are immune to Covid-19, and $98\%$ are still susceptible and face a Covid-19 infection. 
If~the epidemic speeds up now too quickly, the effect can be as in scenario 5 or 6, discussed in this paper. According to the model, initial suppression of an epidemic does not reduce accumulated deaths over the long term, but extends the duration of the epidemic. Most of the European countries decided to suppress the Covid-19 in the first six months after the outbreak. Sweden took a different approach. The comparison \cite{bradde_2020} shows that at the beginning there were relatively more deaths in Sweden than in other European countries, but this comparison does not take into account that the epidemic in Sweden is at a more advanced stage, which means that there are more people who are already immune to Covid-19. One~has to wait with comparisons until the end of the epidemic.

\section{Conclusions}

Let us underline that the model developed in the paper does not attempt to simulate the Covid-19 pandemic but only to imitate some of its aspects. The basic assumptions are that  immunity can be obtained only by infection and that reinfections are rare and can be neglected. Under these assumptions, the pandemic ends only after the herd immunity is achieved. The model has been constructed in a minimalistic way. The scale parameters of the model, like the number of ventilators and mortality rates simulated the real values. The conclusions drawn from the model can be treated qualitatively. Let us recall the main ones:
\begin{itemize}
\item Strong suppression of an epidemic in the early stages does not significantly reduce the total number of deaths over the long term, but increases the duration of the epidemic;
\item In the absence of an efficient drug and a vaccine, the optimal strategy for reducing the total death toll for Covid-19-like epidemics, is to keep the number of new infections at a level where the number of SARS cases is as close as possible to the capacity of the healthcare system.
\item In the early stages of an epidemic, suppression should be only then implemented when one wants to gain time to increase the efficiency of the healthcare system or if the introduction of a drug or a vaccine is expected in a short time.
\end{itemize}
In contrast to the model, in the real world, it is very difficult to fine-tune the parameters that control the rate at which an epidemic spreads and to implement appropriate measures in society, without a vaccine.

\acknowledgments{The author thanks Krzysztof Malarz for interesting discussions and assistance in preparing plots.}



\begin{thebibliography}{999}

\bibitem{Bailey_1975}
N.T. Bailey, {\em The Mathematical Theory of Infectious Diseases}, 2nd ed.,
  (Hafner: New York, NY, USA, 1975).

\bibitem{Anderson_1992}
R.M. Anderson and R.M. May, 
{\em Infectious Diseases of Humans: Dynamics and Control},
(Oxford University Press: Oxford, UK, 1992)

\bibitem{Hethcote_2000}
H.W. Hethcote, {\em The Mathematics of Infectious Diseases}, SIAM Rev. 
{\bf 42},~599--653, (2000).

\bibitem{Li_2018}
M.Y. Li, {\em An Introduction to Mathematical Modeling of Infectious
  Diseases},  (Springer International Publishing AG: Cham, Switzerland, 2018).

\bibitem{Ferguson_2005} N.M. Ferguson, et al.,
{\em Strategies for containing an emerging influenza pandemic in Southeast Asia}, 
Nature {\bf 437},~209--214 (2005).

\bibitem{Ferguson_2020}
N.M. Ferguson, et al.,
{\em Report 9: Impact of non-pharmaceutical interventions (NPIs) to reduce
  COVID-19 mortality and healthcare demand}, (Imperial College COVID-19 Response Team, 2020). 

\bibitem{Flaxman_2020}
S. Flaxman, et al.,
{\em Estimating the effects of non-pharmaceutical interventions on
  COVID-19 in Europe}, Nature {\bf 584},~257--261 (2020).

\bibitem{Bernoulli_1760}
D. Bernoulli,
{\em Essai d'une nouvelle analyse de la mortalit\'{e} caus\'{e}e par
  la petite v\'{e}role et des avantages de l'inoculation pour la pr\'{e}venir},
  (Acad\'{e}mie Royale des Sciences: Paris, France, 1760, pp. 1--45).

\bibitem{Dietz_1988}
K. Dietz, {\em The first epidemic model: A historical note on P. D. En'ko},
Austral. J. Statist. {\bf 30A},~56--65 (1988).

\bibitem{Hamer_1906}
W. Hamer, {\em Epidemic disease in England},
Lancet  {\bf 1},~733--739 (1906).

\bibitem{Ross_1911}
R. Ross, {\em The Prevention of Malaria}, 2nd ed.; (Murray: London, UK, 1911).

\bibitem{Kermack_1927}
W.O. Kermack and A.G. McKendrick, 
{\em A Contribution to the Mathematical Theory of Epidemics},
Proc. R. Soc. Lond. Ser. A {\bf 115},~700--721 (1927).

\bibitem{Pastor-Satorras_2015}
R. Pastor-Satorras, C. Castellano, P. Van~Mieghem and A. Vespignani, 
{\em Epidemic processes in complex networks},
Rev. Mod. Phys. {\bf 87},~925--979 (2015).

\bibitem{Barabasi_1999}
A.L. Barab{\'a}si and R. Albert,
{\em Emergence of Scaling in Random Networks},
Science {\bf 286},~509--512 (1999).

\bibitem{Albert_2002}
R. Albert and A.L. Barab\'asi,
{\em Statistical mechanics of complex networks},
Rev. Mod. Phys. {\bf 74},~47--97 (2002).

\bibitem{Dorogovtsev_2002}
S.N. Dorogovtsev and J.F.F. Mendes, 
{\em Evolution of networks}, Adv. Phys. {\bf 51},~1079--1187 (2002). 

\bibitem{Newman_2003}
M.E.J. Newman, {\em The Structure and Function of Complex Networks},
SIAM Rev. {\bf 45},~167--256 (2002).

\bibitem{Barthelemy_2011}
M. Barth{\'e}lemy,
{\em Spatial networks}, Phys. Rep. {\bf 499},~1--101 (2011).

\bibitem{Chowell_2003}
G. Chowell, J.M. Hyman, J.M., S. Eubank, S. and C. Castillo-Chavez,
{\em Scaling laws for the movement of people between locations in a large
  city}, Phys. Rev. E {\bf 68},~066102 (2003).

\bibitem{Colizza_2006}
V. Colizza, A. Barrat, M. Barth{\'e}lemy and A. Vespignani, 
{\em The role of the airline transportation network in the prediction and
  predictability of global epidemics},
 Proc. Natl. Acad. Sci. USA {\bf 103},~2015--2020 (2006).

\bibitem{Balcan_2009}
D. Balcan, V. Colizza, B. Gon{\c c}alves, H. Hu, J. J. Ramasco, J.J. 
and A. Vespignani,
{\em Multiscale mobility networks and the spatial spreading of infectious
  diseases}, Proc. Natl. Acad. Sci. USA, {\bf 106},~21484--21489 (2009).

\bibitem{Pastor-Satorras_2001}
R. Pastor-Satorras, and A. Vespignani, {\em Epidemic Spreading in Scale-Free Networks},
Phys. Rev. Lett.~{\bf 86},~3200--3203 (2001).

\bibitem{Barrat_2008}
A. Barrat, M. Barth\'{e}lemy and A. Vespignani,
{\em Dynamical Processes on Complex Networks}; (Cambridge University
  Press: Cambridge, UK, 2008).

\bibitem{Miller_2009}
J.C. Miller, {\em Spread of infectious disease through clustered populations},
J. R. Soc. Interface, {\bf 6},~1121--1134 (2009).

\bibitem{Colizza_2007}
V. Colizza, R. Pastor-Satorras and A. Vespignani,
{\em Reaction-diffusion processes and metapopulation models in
  heterogeneous networks}, Nat. Phys. {\bf 3},~276--282 (2007).

\bibitem{Colizza_2008}
V. Colizza and A. Vespignani, 
{\em Epidemic modeling in metapopulation systems with heterogeneous
  coupling pattern: Theory and simulations},
J. Theor. Biol. {\bf 251},~450--467 (2008). 

\bibitem{Bootsma_2007}
M. Bootsma and N.M. Ferguson, 
{\em The effect of public health measures on the 1918 influenza pandemic
  in US cities}, Proc. Natl. Acad. Sci. USA, {\bf 104},~7588--7593 (2007).

\bibitem{Cauchemez_2009}
S. Cauchemez, et al.,
{\em Household Transmission of 2009 Pandemic Influenza A (H1N1) Virus in
  the United States}, N. Engl. J. Med. {\bf 361},~2619--2627 (2009).

\bibitem{Bajardi_2011}
P. Bajardi, et al.,
{\em Human Mobility Networks, Travel Restrictions, and the Global Spread
  of 2009 H1N1 Pandemic}, PLoS ONE {\bf 6},~e16591 (2009).

\bibitem{Otete_2013}
E.H. Otete, et al., 
{\em Parameters for the Mathematical Modelling of Clostridium difficile
  Acquisition and Transmission: A Systematic Review},
PLoS ONE {\bf 8}~e84224 (2013).

\bibitem{Chinazzi_2020} M. Chinazzi, et al.,
{\em The effect of travel restrictions on the spread of the 2019 novel
  coronavirus (COVID-19) outbreak.}
Science {\bf 368},~395--400 (2020).

\bibitem{Kucharski_2020}
A.J. Kucharski, et al., 
{\em Early dynamics of transmission and control of COVID-19: A
  mathematical modelling study},
 Lancet Infect. Dis. {\bf 20},~553--558 (2020).

\bibitem{Zachreson_2019}
C. Zachreson, K.M. Fair, N. Harding and M. Prokopenko,
{\em Interfering with influenza: nonlinear coupling of reactive and static
  mitigation strategies}, J. R. Soc. Interface. {\bf 17}~20190728 (2019).

\bibitem{Li_2020}
Q. Li, et al.,
{\em Early transmission dynamics in Wuhan, China, of novel
  coronavirus-infected pneumonia},
N. Engl. J. Med. {\bf 382},~1199--1207 (2020).

\bibitem{cdc}
Centers for Disease Control and Prevention.  
Available online:  \url{https://www.cdc.gov/} (accessed on September 30, 2020). 

\bibitem{Wu_2020a}
J.T. Wu, K. Leung and G.M. Leung,
{\em Nowcasting and forecasting the potential domestic and international
  spread of the 2019-nCoV outbreak originating in Wuhan, China: a modelling
  study}, Lancet {\bf 395},~689--697 (2020).

\bibitem{Du_2020} Z. Du, et al.,
{\em Risk for transportation of coronavirus disease from Wuhan to other
  cities in China}, Emerg. Infect. Dis. {\bf 26},~1049–-1052 (2020).

\bibitem{Alimohamadi_2020}
Y. Alimohamadi, M. Taghdir and M. Sepandi,
{\em The Estimate of the Basic Reproduction Number for Novel Coronavirus
  disease (COVID-19): A Systematic Review and Meta-Analysis},
J. Prev. Med. Public Health, {\bf 53},~151--157 (2020).

\bibitem{Ledford_2020}
H. Ledford, {\em Coronavirus reinfections: three questions scientists are asking},
Nature {\bf 585},~168--169 (2020).

\bibitem{Wu_2020}
J. T. Wu, et al., {\em Estimating clinical severity of COVID-19 from the transmission
  dynamics in Wuhan, China}, Nat. Med. {\bf 26},~506--510 (2020).

\bibitem{VPort}
R. R. Nunes, {\em Covid-19. Apenas 5\% dos doentes podem precisar de um ventilador},
Dia\'ario de Noticias, March 2020, Available online: 
\url{https://www.dn.pt/pais/covid-19-apenas-5-dos-doentes-podem-precisar-de-um-ventilador-11949111.html} (accessed on September 30, 2020).

\bibitem{VR}
I. Golunov, A. Kovalev and D. Sarkisyan, 
{\em The ventilator problem Russia has way more machines that can keep
  coronavirus patients breathing than Italy did---but that doesn’t mean the
  pandemic will be any easier},
Meduza March 20, 2020, { Available online: 
  \url{https://meduza.io/en/feature/2020/03/21/the-ventilator-problem}} (accessed on September 30, 2020).

\bibitem{VG}
S. Kliff, A. Satariano, J. Silver-Greenberg and N. Kulish,
{\em There Aren't Enough Ventilators to Cope With the Coronavirus},
The New York Times, March 18, 2020,
Available online: 
\url{https://www.nytimes.com/2020/03/18/business/coronavirus-ventilator-shortage.html}
  (accessed on September 30, 2020).

\bibitem{VP}
L. Szyma\'nski, 
{\em Poland Has Time for Action---Minister on Coronavirus},
The First News, PAP,  May 10, 2020, Polish Press Agency, Warsaw, Poland,
Available online: https://www.thefirstnews.com/article/poland-has-time-for-action---minister-on-coronavirus-11018, (accessed September 30,2020).

\bibitem{who}
World Health Organization. Available online: \url{https://www.who.int/healthinfo/mortality\_data/en/} (accessed on 
September 30, 2020).

\bibitem{Richards2020}
M. Richards, M. Anderson, P. Carter, B.L. Ebert and E. Mossialos,
{\em The impact of the COVID-19 pandemic on cancer care},
Nat. Cancer {\bf 1},~565--567 (2020).

\bibitem{Maringe2020}
C. Maringe, et al., {\em The~impact of the COVID-19 pandemic on cancer deaths due to delays in diagnosis in England, UK: a national, population-based, modelling study},
Lancet Oncol. {\bf 21},~1023--1034 (2020).

\bibitem{Pessoa-Amorim}
G. Pessoa-Amorim, et al., {\em Admission of patients with STEMI since the outbreak of the COVID-19 pandemic: A survey by the European Society of Cardiology},
{Eur.~Heart J. Qual. Care Clin. Outcomes}, {\bf 6},~210--216 (2020).

\bibitem{PP}
A. Sowa, {\em Rak przegrywa z wirusem. Tysi\c{a}ce os\'{o}b bez diagnozy},
Polityka, June 10, 2020.
{Available~online: \url{https://www.polityka.pl/tygodnikpolityka/spoleczenstwo/1959930,1,rak-przegrywa-z-wirusem-tysiace-osob-bez-diagnozy.read}} (accessed on September 30, 2020).

\bibitem{Dall_2002}
J. Dall and M. Christensen, 
{\em Random Geometric Graphs}, Phys. Rev. {\bf E 66},~016121 (2002).

\bibitem{Penrose_2003}
M. Penrose, {\em Random Geometric Graphs}, (Oxford University Press, Oxford, UK, 2003).

\bibitem{Erdos_1959}
Erd{\H o}s, P.; R\'enyi, A.
{\em On random graphs.}, Publ. Math. {\bf 6}, 290--297, (1959).

\bibitem{Moore_2020}
S. Moore and T. Rogers,
{\em Predicting the Speed of Epidemics Spreading in Networks},
Phys. Rev. Lett. {\bf 124}, 068301, (2020).

\bibitem{krn}
National Cancer Registry in Poland., Available~online: \url{http://onkologia.org.pl/} (accessed on September 30, 2020).

\bibitem{who_covid}
Coronavirus Disease (COVID-19) Weekly Epidemiological Update and Weekly
Operational Update as of 27 September 2020. Available~online: \url{https://www.who.int/emergencies/diseases/novel-coronavirus-2019/situation-reports}
  (accessed on September 30, 2020).

\bibitem{bradde_2020}
S. Bradde, B. Cerruti and J.-P. Bouchaud, 
{\em Did lockdowns serve their purpose?}, arXiv:physics.soc-ph/2006.09829 (2020).

\end{thebibliography}
\end{document}